\documentclass[onecolumn,showpacs,amsmath,amssymb,nofootinbib,12pt]{revtex4-1}
\usepackage{graphicx}
\usepackage{dcolumn}
\usepackage{bm}
\begin{document}
\newcommand{\hs}{\hspace*{0.5cm}}
\newcommand{\vs}{\vspace*{0.5cm}}
\newcommand{\be}{\begin{equation}}
\newcommand{\ee}{\end{equation}}
\newcommand{\bea}{\begin{eqnarray}}
\newcommand{\eea}{\end{eqnarray}}
\newcommand{\ben}{\begin{enumerate}}
\newcommand{\een}{\end{enumerate}}
\newcommand{\bde}{\begin{widetext}}
\newcommand{\ede}{\end{widetext}}
\newcommand{\nn}{\nonumber}
\newcommand{\crn}{\nonumber \\}
\newcommand{\Tr}{\mathrm{Tr}}
\newcommand{\non}{\nonumber}
\newcommand{\noi}{\noindent}
\newcommand{\al}{\alpha}
\newcommand{\la}{\lambda}
\newcommand{\bet}{\beta}
\newcommand{\ga}{\gamma}
\newcommand{\va}{\varphi}
\newcommand{\om}{\omega}
\newcommand{\pa}{\partial}
\newcommand{\+}{\dagger}
\newcommand{\fr}{\frac}
\newcommand{\bc}{\begin{center}}
\newcommand{\ec}{\end{center}}
\newcommand{\Ga}{\Gamma}
\newcommand{\de}{\delta}
\newcommand{\De}{\Delta}
\newcommand{\ep}{\epsilon}
\newcommand{\varep}{\varepsilon}
\newcommand{\ka}{\kappa}
\newcommand{\La}{\Lambda}
\newcommand{\si}{\sigma}
\newcommand{\Si}{\Sigma}
\newcommand{\ta}{\tau}
\newcommand{\up}{\upsilon}
\newcommand{\Up}{\Upsilon}
\newcommand{\ze}{\zeta}
\newcommand{\ps}{\psi}
\newcommand{\Ps}{\Psi}
\newcommand{\ph}{\phi}
\newcommand{\vph}{\varphi}
\newcommand{\Ph}{\Phi}
\newcommand{\Om}{\Omega}
\newcommand{\AdrHEPC}{$^1$Phenikaa Institute for Advanced Study, Phenikaa University, Hanoi 100000, Vietnam\\
$^2$Faculty of Basic Science and Faculty of Materials Science and Engineering,\\ Phenikaa University, Hanoi 100000, Vietnam}

\title{Phenomenology of the simple 3-3-1 model with inert scalars} 

\author{Phung Van Dong$^{1,2,*}$, N. T. K. Ngan$^3$, T. D. Tham$^4$, L. D. Thien$^5$, and N. T. Thuy$^6$}
\affiliation{\AdrHEPC\\ 
$^3$Department of Physics, Cantho University, 3/2 Street, Ninh Kieu, Cantho, Vietnam\\
$^4$Faculty of Natural Science Education, Pham Van Dong University, 509 Phan Dinh Phung, Quang Ngai City, Vietnam\\
$^5$Graduate University of Science and Technology, Vietnam Academy of Science and Technology, 18 Hoang Quoc Viet, Cau Giay, Hanoi, Vietnam\\
$^6$Science Department, Bac Ninh Specialized High School, 4 Ho Ngoc Lan, Bac Ninh City, Vietnam\\
$^*$Email: dong.phungvan@phenikaa-uni.edu.vn} 

\date{\today}

\begin{abstract}
The simple 3-3-1 model that contains the minimal lepton and minimal scalar contents is detailedly studied. The impact of the inert scalars (i.e., the extra fundamental fields that provide realistic dark matter candidates) on the model is discussed. All the interactions of the model are derived, in which the standard model ones are identified. We constrain the standard model like Higgs particle at the LHC. We search for the new particles including the inert ones, which contribute to the $B_s$-$\bar{B}_s$ mixing, the rare $B_s\rightarrow \mu^+\mu^-$ decay, the CKM unitarity violation, as well as producing the dilepton, dijet, diboson, diphoton, and monojet final states at the LHC.            

\end{abstract}


\maketitle

\section{\label{intro}Introduction}

One of the promising extensions of the standard model is the model based upon the $SU(3)_C\otimes SU(3)_L\otimes U(1)_X$ (3-3-1) gauge symmetry \cite{331m,331r}. Indeed, it can explain neutrino masses~\cite{neutrino331}, dark matter \cite{dm331,dns,3311}, fermion generation number \cite{gen}, oddly heavy top quark~\cite{tquark}, flavor physics~\cite{flav331}, electric charge quantization \cite{ecq}, and strong CP conservation~\cite{palp}\footnote{On the other hand, this framework can be naturally unified with left-right symmetry which addresses both the weak parity violation and fermion generation number, besides the implication for neutrino mass generation and dark matter stability \cite{331lradd}.}. Along this evolution, the numerous variants of the 3-3-1 model have been proposed so far, but they mainly differ in the lepton and scalar contents. Given a favor of the version with minimal lepton and scalar contents, the suitable one is the so-called simple 3-3-1 model. This model was first introduced in \cite{dns} in order to resolve the problems associated with the reduced (minimal) 3-3-1 model \cite{r331}. Additionally, dark mater might be naturally implemented in this new proposal, as also stated in \cite{dns}.    

It was shown in \cite{ds} that the considering model encounters a serious discrepancy between the flavor-changing neutral current (FCNC) and $\rho$-parameter bounds on the new physics scale, which was experimentally unacceptable. This could be understood as follows. The FCNC constrains the 3-3-1 breaking scale to be $w>3.6$ TeV. With this limit, the new physics contribution of the normal sector of the simple 3-3-1 model to the $\rho$-parameter is negligible. This implies that the inert scalars are necessarily included as they are, which provide dark matter candidates. The presence of the inert scalars can solve the experimental $\rho$-parameter, $(\Delta\rho)_{\mathrm{new-physics}}=0.00039\pm 0.00019$, which is 2 $\sigma$ above the standard model prediction, $\rho=1$ \cite{pdg}. Indeed, one of the inert doublets contained in the inert multiplet $\phi=\eta'$, $\chi'$ or $\sigma$ can be used to break the vector part of weak $SU(2)$ and lead to a positive contribution such as $(\Delta \rho)_{\mathrm{inert-doublet}} = \fr{G_{F}}{8\sqrt{2}\pi^2} \Delta m^2$, where  
\be \Delta m^2= m^2_1
+m^2_2-\fr{4m^2_1 m^2_2}{m^2_1-m^2_2}\ln\fr{m_1}{m_2}\ee denotes the mass splitting between the inert doublet components \cite{pdg}. Compared to the bound, it yields $(27.7\ \mathrm{GeV})^2<\Delta m^2<(83\ \mathrm{GeV})^2$ at 90\% CL, in good agreement with the dark matter constraints where $m_1$ and $m_2$ can range from the weak scale to TeV scale~\cite{dns}.

The above observation makes the simple 3-3-1 model realistic (i.e., alive) as well as showing the importance of the inert scalars, besides the other strong connections between the two sectors as presented in \cite{dns}. Motivated by this fact, in the current work, the normal sector of the simple 3-3-1 model is investigated in detail, taking into account the effect of the inert sector too. It is noted that the inert sector was explicitly obtained in \cite{dns}, which would be appropriately used in this work whenever it contributes to the normal sector as well as the physical processes of interest. 

First of all, we give a brief review of the simple 3-3-1 model with inert scalars, where all the interactions in the normal sector will be calculated. The interactions between the normal and inert sectors were available, which could be found in \cite{dns}. Next, we consider the standard model like Higgs particle signature at the LHC when additionally including the effects of the new particles. We also examine the decay anomaly $B_s\rightarrow \mu^+\mu^-$, the $B_s$-$\bar{B}_s$ mixing, and the violation of CKM unitarity, which constrain $Z'$ boson. The searches for the new scalars $H^0$, $H^\pm$, $\phi$ and the new gauge bosons $Z'$, $X^\pm$, $Y^{\pm\pm}$ are lastly discussed.

Thus, the rest of this work is organized as follows. In Sec. \ref{model}, the necessary features of the model are given. In Sec. \ref{interaction}, we calculate relevant interactions. Section \ref{pheno} constrains the standard model like Higgs particle and searches for the new physics. Finally, we summarize our results and make conclusions in Sec. \ref{conclusion}.    

\section{\label{model} The model}

The gauge symmetry of the model is given by \be SU(3)_C\otimes SU(3)_L\otimes U(1)_X,\ee where the first factor is the ordinary QCD group, while the last two are a nontrivial extension of the electroweak group. 

The standard model fermion doublets will be enlarged to $SU(3)_L$ triplets/antitriplets, i.e. $3=2\oplus 1$ and/or $3^*=2^*\oplus 1$ with $2^*=i\sigma_2 2$, whereas the standard model fermion singlets, possibly including right-handed neutrinos, are either retained as corresponding $SU(3)_L$ singlets or included along with the fermion doublets to form the mentioned triplets/antitriplets. The latter case may be valid for leptons, but it does not apply for quarks because of the commutation among $SU(3)_C$, $SU(3)_L$, and the spacetime symmetry. Therefore, the introduction of exotic quarks to complete the quark triplets/antitriplets is necessary. 

Such arrangement of fermions hints the fermion generation number to be equal the fundamental color number (i.e., 3), when the anomaly cancelation and the QCD asymptotic freedom are imposed. The above latter case, i.e. the lepton triplets/antitriplets that include right-handed charged leptons or neutrinos, implies the quantization of electric charge, when the anomaly cancelation and the mass generation are applied. 

In the following, we are interested in the model whose lepton sector uses only those of the standard model, i.e. no new lepton is required, and of course the electric charge is embedded in the gauge symmetry as \be Q=T_3-\sqrt{3}T_8+X,\ee where $T_i$ ($i=1,2,3,..,8$) and $X$ are the generators of $SU(3)_L$ and $U(1)_X$, respectively. Furthermore, the $SU(3)_C$ generators will be denoted as $t_i$. 

In summary, the fermion content which is anomaly free is given by 
\bea \psi_{aL} &\equiv & \left(\begin{array}{c}
               \nu_{aL}  \\ e_{aL} \\ (e_{aR})^c
\end{array}\right) \sim (1,3,0),\\
Q_{\al L}  &\equiv& \left(\begin{array}{c}
  d_{\al L}\\  -u_{\al L}\\  J_{\al L}
\end{array}\right)\sim (3,3^*,-1/3),\\ 
Q_{3L} &\equiv& \left(\begin{array}{c} u_{3L}\\  d_{3L}\\ J_{3L} \end{array}\right)\sim
 \left(3,3,2/3\right), \\ u_{a
R}&\sim&\left(3,1,2/3\right),\hs d_{a R} \sim \left(3,1,-1/3\right),\\
J_{\al R} &\sim&
\left(3,1,-4/3\right),\hs J_{3R} \sim \left(3,1,5/3\right), \eea where $a=1,2,3$ and $\al=1,2$ are generation indices. The quantum numbers in the parentheses $(N_C,N_L,X)$ define corresponding representations under the $(SU(3)_C, SU(3)_L, U(1)_X)$ groups, respectively. The superscript $^c$ indicates the charge conjugation, $(e_R)^c\equiv C\bar{e}^T_R=(e^c)_L$, as usual. The new quarks $J_a$ have exotic electric charges such as $Q(J_\al)=-4/3$ and $Q(J_3)=5/3$, whereas the other fermions ($\nu,e,u,d$) possess usual electric charges. 

Notably, the $[SU(3)_L]^3$ anomaly cancellation requires  one of quark generations transforming differently from the remaining quark generations. Here, the third generation of quarks has been arranged differently from the first two, in order to have a well-defined new physics scale below the Landau pole of around 5 TeV, due to the constraints of the FCNCs \cite{dns}.      

To break the gauge symmetry and generate the masses for particles in a correct way, the scalar content can minimally be introduced as  
\bea \eta = \left(\begin{array}{c}
\eta^0_1\\
\eta^-_2\\
\eta^{+}_3\end{array}\right)\sim (1,3,0),\hs \chi = \left(\begin{array}{c}
\chi^-_1\\
\chi^{--}_2\\
\chi^0_3\end{array}\right)\sim (1,3,-1),\label{vev2}
\eea with corresponding vacuum expectation values (VEVs), \bea \langle \eta\rangle = \fr{1}{\sqrt{2}}
\left(\begin{array}{c}
u\\ 0\\ 0\end{array}\right),\hs  \langle \chi\rangle =\fr{1}{\sqrt{2}}
\left(\begin{array}{c}
0\\ 0\\ w\end{array}\right).\eea 

This scalar sector is unique, given simultaneously that (a) it includes only two scalar triplets, (b) the top quark gets a tree-level mass, and (c) the $\rho$-parameter coincides with the global fit, aforementioned. Let us stress that another choice of two scalar triplets such as $\rho$ and $\chi$ as in~\cite{r331} would lead to an unacceptably large contribution for the $\rho$-parameter~\cite{ds}; additionally, such choice yields a vanishing tree-level mass for the top quark, which is unnaturally induced by the radiative corrections or effective interactions \cite{dns}. 

The VEV $w$ breaks the 3-3-1 symmetry down to the standard model symmetry and gives the masses for the new particles, while $u$ breaks the standard model symmetry down to $SU(3)_C\otimes U(1)_Q$ and provides the masses for the ordinary particles. To keep consistency with the standard model, we impose $u\ll w$.    

The dark sector is denoted by $\phi$ that is an inert scalar multiplet, $\phi=\eta',\ \chi'$, or $\sigma$, ensured by an extra $Z_2$ symmetry, $\phi\rightarrow -\phi$, which provides dark matter candidates \cite{dns}. Interestingly enough, the presence of $\phi$ is crucial to have a realistic simple 3-3-1 model, not only due to the dark matter solution. Indeed, it provides $B-L$ violating interactions, that along with the following one responsible for neutrino mass generation realize $B-L$ as an approximate symmetry, which is acquired in this kind of the models. Otherwise, if $B-L$ is an exact symmetry, the model is not self-consistent because the $B-L$ and 3-3-1 symmetries are algebraically non-closed \cite{3311,d3311}. Furthermore, those interactions also separate the inert field masses which make the dark matter candidates viable \cite{dns}. Lastly, $\phi$ governs the $\rho$-parameter by which it is comparable with the global fit, as mentioned. In other words, the $\rho$-parameter can contain information on the dark sector.          

The total Lagrangian, up to the gauge fixing and ghost terms, is obtained by 
\bea \mathcal{L} &=& \sum_F \bar{F}i\ga^\mu D_\mu F + \sum_S (D^\mu S)^\dagger (D_\mu S) \crn
&& -\fr 1 4 G_{i\mu\nu} G^{\mu\nu}_i-\fr 1 4 A_{i \mu\nu} A^{\mu\nu}_i -\fr 1 4 B_{\mu\nu}B^{\mu\nu}\crn
&&+ \mathcal{L}_{Y}-V,\eea where $F$ and $S$ run over all the fermion multiplets and scalar multiplets, respectively. The covariant derivative is $D_\mu=\pa_\mu + i g_s t_i G_{i\mu} + i g T_i A_{i\mu} + i g_X X B_\mu$, where $(g_s,g,g_X)$ and $(G_{i\mu},A_{i\mu},B_\mu)$ are the gauge couplings and gauge bosons of the 3-3-1 groups, respectively. The field strength tensors $G_{i\mu\nu}$, $A_{i\mu\nu}$, and $B_{\mu\nu}$ are correspondingly followed. 

The scalar potential is given by $V=V_{\mathrm{simple}}+V_{\mathrm{inert}}$, where the first term is 
\be V_{\mathrm{simple}}=\mu^2_1\eta^\dagger \eta + \mu^2_2 \chi^\dagger \chi + \la_1 (\eta^\dagger \eta)^2+\la_2 (\chi^\dagger \chi)^2+\la_3(\eta^\dagger\eta)(\chi^\dagger \chi) +\la_4(\eta^\dagger \chi)(\chi^\dagger\eta),\ee which is the potential of the normal scalar sector ($\eta,\chi$). Whereas, the second term $V_{\mathrm{inert}}$ that includes the potential of the inert scalar sector ($\phi$) as well as the interactions between the two (inert, normal) sectors is identical to those supplied in \cite{dns} for $\phi=\eta'$, $\chi'$, or $\sigma$ respectively, which should be kept for brevity. 

Lastly, the Yukawa Lagrangian takes the form, \bea \mathcal{L}_Y&=&h^J_{33}\bar{Q}_{3L}\chi J_{3R}+ h^J_{\al \beta}\bar{Q}_{\al L} \chi^* J_{\beta R}+h^u_{3 a} \bar{Q}_{3L} \eta u_{aR}+ \fr{h^u_{\al a}}{\La} \bar{Q}_{\al L} \eta \chi u_{a R}\crn
&&+ h^d_{\al a} \bar{Q}_{\al L} \eta^* d_{a R} + \fr{h^d_{3 a}}{\La} \bar{Q}_{3L}\eta^*\chi^* d_{a R} +h^e_{ab} \bar{\psi}^c_{aL} \psi_{bL}\eta + \fr{h'^e_{ab}}{\La^2}(\bar{\psi}^c_{aL}\eta\chi)(\psi_{bL}\chi^*)\crn 
&&+\fr{s^\nu_{ab}}{\La} (\bar{\psi}^c_{aL}\eta^*)(\psi_{bL} \eta^*)+H.c.,\eea where $\La$ is a new scale that defines the effective interactions, needed to generate appropriate masses for all the fermions \cite{dns}. All the $h$'s couplings conserve $B-L$, except for the $s^\nu$ coupling that violates this charge by two units. This reflects the fact $B-L$ as of an approximate symmetry, and that the corresponding neutrino masses derived are reasonably small \cite{dns}.       

Because of the $Z_2$ symmetry, the VEV of $\phi$ vanishes, $\langle \phi \rangle=0$. The gauge bosons gain masses from the VEVs of $\eta$ and $\chi$, given by the Lagrangian $\sum_{S=\eta,\chi}(D^\mu\langle S\rangle)^\dagger (D_\mu\langle S\rangle)$. The gluons $G_{i}$ have zero mass, as usual. The physical charged gauge bosons with corresponding masses are obtained by 
\bea &&W^{\pm}= \fr{A_1\mp i A_2}{\sqrt{2}},\hs m^2_W=\fr{g^2}{4}u^2,\\
&&X^\mp = \fr{A_4 \mp i A_5}{\sqrt{2}},\hs m^2_X=\fr{g^2}{4}(w^2+u^2),\\
&& Y^{\mp\mp}= \fr{A_6 \mp i A_7}{\sqrt{2}},\hs m^2_Y=\fr{g^2}{4}w^2. \eea 
The $W$ boson is identical to that of the standard model, which yields $u\simeq 246\ \mathrm{GeV}$. $X$ and $Y$ are new charged gauge bosons which have large masses in the $w$ scale, due to $w\gg u$. 

The neutral gauge bosons with corresponding masses are achieved as \cite{dns}
\bea && A = s_W A_{3}+ c_W\left(-\sqrt{3}t_W A_{8}+\sqrt{1-3t^2_W}B\right),\hs m_A=0,\\
&& Z= c_W A_{3}- s_W\left(-\sqrt{3}t_W A_{8}+\sqrt{1-3t^2_W}B\right),\hs m^2_{Z} = \fr{g^2}{4c^2_W}u^2,\\
&& Z'=\sqrt{1-3t^2_W}A_{8}+\sqrt{3}t_W B,\hs m^2_{Z'}=\fr{g^2[(1-4s^2_W)^2u^2+4c^4_W w^2]}{12 c^2_W(1-4s^2_W)},
\eea
where $s_W=e/g = t/\sqrt{1+4t^2}$, with $t=g_X/g$, is the sine of the Weinberg angle \cite{dl}.
 
Note that the photon field $A$ is massless and decoupled, i.e. a physical particle, whereas $Z$ and $Z'$ slightly mix via a mass term, given by $m^2_{ZZ'}=\fr{g^2\sqrt{1-4s^2_W}}{4\sqrt{3}c^2_W}u^2\simeq 0.16 m^2_Z\ll m^2_{Z'}$, with the aid of $s^2_W\simeq 0.231$. The mixing angle of $Z$-$Z'$ is defined by 
$t_{2\varphi}=2m^2_{ZZ'}/(m^2_{Z'}-m^2_Z)\simeq  1.4\times 10^{-4} \times \left(\fr{3.6\ \mathrm{TeV}}{w}\right)^2$. The $Z$-$Z'$ mixing term leads to the shifts in $Z$, $Z'$ masses, determined by $\Delta m^2_Z/m^2_Z \simeq  -1.14\times 10^{-5}\times \left(\fr{3.6\ \mathrm{TeV}}{w}\right)^2$ and $\Delta m^2_{Z'}/m^2_{Z'} \simeq 5.1\times 10^{-9}\times \left(\fr{3.6\ \mathrm{TeV}}{w}\right)^4$, and the deviation of the $\rho$ parameter is $(\Delta \rho)_{\mathrm{mixing}} \simeq -\Delta m^2_Z/m^2_Z$. All such mixing effects are infinitesimal, which can be neglected, due to $w>3.6$ TeV from the FCNC bound \cite{dns}. The $Z$ boson is a physical particle, identical to that of the standard model, while $Z'$ is a new neutral gauge boson with a large mass in the $w$ scale. 

The experimental $\rho$ parameter can be explained by the loop effect of the inert scalar $\phi$, as shown in the introduction. Let us remind the reader that the loop effect of $X$, $Y$ gauge bosons is negligible \cite{ds}, similar to the above mixing effect.    

Because of the $Z_2$ symmetry, the normal scalars do not mix with the inert scalars. Also, the physical eigenstates and masses of the normal scalars are given from $V_{\mathrm{simple}}$, which are distinct from the inert sector (see also Dong {\it et al.} in \cite{dm331}). Let us expand   
\bea \eta=\left(\begin{array}{c} \frac{u}{\sqrt{2}} \\ 0 \\ 0 \end{array}\right) + \left(\begin{array}{c} \frac{S_1+iA_1}{\sqrt{2}} \\ \eta_2^- \\ \eta_3^+ \end{array} \right), \hs \chi=\left(\begin{array}{c} 0 \\ 0 \\ \frac{w}{\sqrt{2}}\end{array}\right) + \left(\begin{array}{c} \chi_1^- \\ \chi_2^{--} \\ \frac{S_3+iA_3}{\sqrt{2}} \end{array}\right),\label{VEVs}\eea which yields the physical Higgs particles with corresponding masses,  
\bea 
&& h\equiv c_\xi S_1-s_\xi S_3,\ \ m^2_h=\la_1 u^2+\la_2 w^2-\sqrt{(\la_1 u^2-\la_2 w^2)^2+\la^2_3 u^2 w^2}\simeq \fr{4\la_1\la_2-\la^2_3}{2\la_2}u^2,\crn
&& H\equiv s_\xi S_1 + c_\xi S_3,\hs m^2_{H}=\la_1 u^2+\la_2 w^2+\sqrt{(\la_1 u^2-\la_2 w^2)^2+\la^2_3 u^2 w^2}\simeq 2\la_2 w^2,\\
&& H^{\pm}\equiv c_{\theta}\eta^\pm_3+s_{\theta}\chi^\pm_1,\hs m^2_{H^\pm}=\fr{\la_4}{2}(u^2+w^2),\nn \eea where 
$\xi$ is the mixing angle of $S_1$-$S_3$, while $\theta$ is that of $\chi_1$-$\eta_3$, obtained as $ t_\theta=u/w,\ t_{2\xi}=\la_3 u w/(\la_2 w^2-\la_1 u^2)\simeq (\la_3 u)/(\la_2 w)$. The fields $A_1$, $A_3$ are massless Goldstone bosons eaten by $Z$, $Z'$, respectively, $G_{Z}\equiv A_1,  G_{Z'}\equiv A_3$. The fields $\chi_2^{\pm\pm}$, $\eta_2^\pm$ are massless Goldstone bosons eaten by $Y^{\pm\pm}$, $W^\pm$, respectively, $G_Y^{\pm\pm}\equiv \chi_2^{\pm\pm},\ G_W^{\pm}\equiv \eta_2^\pm$. The field that is orthogonal to $H^\pm$, $G_X^{\pm} = c_\theta\chi_1^\pm - s_\theta\eta_3^\pm$, is massless Goldstone boson of $X^\pm$. 

In summary, we have four massive Higgs bosons ($h$, $H$, $H^\pm$), in which $h$ is the standard model like Higgs particle (verified below) with a light mass in the $u$ scale, while the others are new Higgs bosons with large masses in the $w$ scale. There are eight Goldstone bosons ($G_Z$, $G_{Z'}$, $G_W^{\pm}$, $G_Y^{\pm\pm}$, and $G_X^{\pm}$) as eaten by the corresponding eight massive gauge bosons. At the effective limit, $u\ll w$, it follows  
\bea \eta\simeq \left(\begin{array}{c} \frac{u+h+iG_Z}{\sqrt{2}} \\ G_W^- \\ H^+ \end{array} \right), \hs \chi\simeq \left(\begin{array}{c} G_X^- \\ G_Y^{--} \\ \frac{w+H+iG_{Z'}}{\sqrt{2}} \end{array}\right).\eea

Finally, let us remind the reader that the physical eigenstates and masses of the inert scalars are derived from $V_{\mathrm{inert}}$ when $\eta$ and $\chi$ develop (replaced by) the VEVs. They were explicitly given in \cite{dns}. The conditions on the scalar potential parameters, which obey simultaneously that (a) the potential is bounded from below, (b) the VEVs $u,w$ are definitely nonzero, (c) the physical scalar masses are definitely positive, and (d) the $Z_2$ symmetry is unbroken by the vacuum (i.e. $\langle \phi\rangle =0$), were also achieved therein.

\section{\label{interaction} Interaction} 
\subsection{Interactions of fermions with gauge bosons} 

The relevant interactions arise from $\sum_F \bar{F}i\ga^\mu D_\mu F$, in which we separate $D_{\mu}= \partial_{\mu}+ig_st_iG_{i\mu}+igP^{\mathrm{CC}}_{\mu}+
igP^{\mathrm{NC}}_{\mu}$, where $P^{\mathrm{CC}}_\mu=\sum_{i\neq 3,8} T_i A_{i\mu}$ and $P^{\mathrm{NC}}_{\mu} =T_3 A_{3\mu}+T_8 A_{8\mu}+t X B_\mu$. The last two terms in $D_\mu$ will produce the charged and neutral currents respectively discussed in this section. Note that since $T_i(F_R)=0$, the charged current includes only left-handed fermions, while the neutral current contains both left- and right-handed fermions. 

\subsubsection{\label{CC} Charged current}

Let us work in the weak basis consisting of the weight-raising and weight-lowering operators, defined by 
\be T_\pm =\fr{T_1\pm iT_2}{\sqrt{2}},\hs U_\pm =\fr{T_4\pm iT_5}{\sqrt{2}},\hs V_\pm =\fr{T_6\pm iT_7}{\sqrt{2}}.\ee 
The corresponding gauge bosons are 
\be W^\pm =\fr{A_1\mp i A_2}{\sqrt{2}},\hs X^\mp =\fr{A_4\mp i A_5}{\sqrt{2}},\hs Y^{\mp\mp} =\fr{A_6\mp i A_7}{\sqrt{2}}, \ee
such that \be P^{\mathrm{CC}}_{\mu} =T_+ W^+_\mu+U_+ X^-_\mu + V_+ Y^{--}_\mu+H.c.\ee Here the superscripts on the fields indicate the electric charges, while the subscripts on the operators are simply marks\footnote{In the literature, the operators are sometimes specified by the corresponding root vectors.}, and we have  $T_-=(T_+)^\dagger$ and so forth for $U$, $V$. 

The charged current takes the form,  
\bea
-g\sum_{F}
\bar{F}\gamma^\mu P_{\mu}^{\mathrm{CC}}
F  = -gJ^\mu_W W^+_\mu-gJ^\mu_X X^{-}_\mu-gJ^\mu_Y Y^{--}_\mu+H.c.,\eea
where
\bea
&& J_W^{\mu}=\sum_F \bar{F}\ga^\mu T_+ F = \frac{1}{\sqrt{2}}\left(\bar{\nu}_{aL}\ga^{\mu}e_{aL}+\bar{u}_{aL}\ga^{\mu}d_{aL}\right),\\
&& J_X^{\mu}=\sum_F \bar{F}\ga^\mu U_+ F=\frac{1}{\sqrt{2}}\left( \bar{\nu}_{aL}\ga^{\mu} e^c_{aR}-\bar{J}_{\al L}\ga^{\mu}d_{\al L}+\bar{u}_{3L}\ga^{\mu}J_{3L}\right),\label{cccd}\\
&& J_Y^{\mu}=\sum_F\bar{F}\ga^\mu V_+ F =\frac{1}{\sqrt{2}}\left(\bar{e}_{aL}\ga^{\mu}e^c_{aR}
+\bar{J}_{\al L}\ga^{\mu}u_{\al L}+\bar{d}_{3L}\ga^{\mu}J_{3L} \right).\eea

\subsubsection{\label{neutral current} Neutral current}
Substituting $A_{3\mu},\ A_{8\mu},\ B_{\mu}$ in terms of $A_{\mu},\ Z_\mu,\ Z'_\mu$ into $P^{\mathrm{NC}}_\mu$, we obtain
\be P_\mu^{\mathrm{NC}}= s_WQA_\mu  +\frac{1}{c_W} \left(T_3-s_W^2 Q\right) Z_{\mu}+\frac{1}{c_W}\left(\sqrt{1-4s^2_W} T_8+\frac{\sqrt{3}s^2_W}{\sqrt{1-4s^2_W}}X \right)Z'_{\mu}. \label{PNC}\ee
The neutral current takes the form,
\bea 
 -g\sum_F\bar{F}\ga^\mu P_\mu^{\mathrm{NC}}F &=& -eQ(f)\bar{f}\gamma^\mu f A_\mu -\frac{g}{2c_W}\bar{f}\gamma^\mu\left[g_V^{Z}(f)-g_A^{Z}(f)\gamma_5\right]fZ_{\mu}\crn
 &&-\frac{g}{2c_W}\bar{f}\gamma^\mu\left[g_V^{Z'}(f)-g_A^{Z'}(f)\gamma_5\right]fZ'_{\mu},\label{fin1} \eea
where $f$ indicates all the fermions of the model, and 
\bea 
&& g_V^Z(f) = T_3(f_L)-2s^2_W Q(f),\hs g_A^{Z}(f) =  T_3(f_L),\\
&& g_V^{Z'}(f) = \sqrt{1-4s^2_W} T_8(f_L)+\frac{\sqrt{3} s^2_W}{\sqrt{1-4s^2_W}}(X+Q)(f_L),\\
&& g_A^{Z'}(f)= \fr{c^2_W}{\sqrt{1-4s^2_W}} T_8(f_L)-\fr{\sqrt{3}s^2_W}{\sqrt{1-4s^2_W}} T_3(f_L).
\eea 

The values of $g_V(f)$ and $g_A(f)$ corresponding to $Z$ and $Z'$ are listed in Tables \ref{tab1} and \ref{tab2}, respectively.  

\begin{table}[h]
\begin{tabular}{|c|c|c|}
\hline
$f$ &  $g^Z_V$ &  $g^Z_A$ \\ \hline 
$\nu_e,\nu_\mu,\nu_\tau$ &  $\frac{1}{2}$&
$\frac{1}{2}$\\ 
$e,\mu,\tau$ & 
$\frac{1}{2}\left(4s^2_W-1\right)$  &
$-\frac{1}{2}$\\ 
$u,c,t$ & $\frac{1}{2}\left(1-\frac{8}{3}s^2_W\right)$ & $\frac{1}{2}$\\
$d,s,b$ & $\frac{1}{2}\left(\frac{4}{3}s^2_W-1\right)$ & $-\frac{1}{2}$\\
$J_1, J_2$  & $\frac{8}{3}s^2_W$ & 0\\
$J_3$ & $-\frac{10}{3}s^2_W$ &
$0$\\ 
\hline 
\end{tabular}
\caption{\label{tab1} The
couplings of $Z$ with fermions.}
\end{table}

\begin{table}[h]
\begin{tabular}{|c|c|c|}
\hline
$f$ &  $g^{Z'}_V$ &  $g^{Z'}_A$ \\ \hline 
$\nu_e,\nu_\mu,\nu_\tau$ & $\frac{1}{2}\sqrt{\frac{1-4s^2_W}{3}}$&
$\frac{1}{2}\sqrt{\frac{1-4s^2_W}{3}}$\\ 
$e,\mu,\tau$ &
$\frac{\sqrt{3}}{2}\sqrt{1-4s^2_W}$ &
$-\frac{1}{2}\sqrt{\frac{1-4s^2_W}{3}}$\\
$u, c$ & $-\frac{1}{2}\frac{1-6s^2_W}{\sqrt{3\left(1-4s^2_W\right)}}$ & $-\frac{1}{2}\frac{1+2s^2_W}{\sqrt{3\left(1-4s^2_W\right)}}$\\
$t$ & $\frac{1}{2}\frac{1+4s^2_W}{\sqrt{3\left(1-4s^2_W\right)}}$ &
$\frac{1}{2}\sqrt{\frac{1-4s^2_W}{3}}$\\ 
$d, s$ & $-\frac{1}{2}\frac{1}{\sqrt{3\left(1-4s^2_W\right)}}$ & $-\frac{1}{2}\sqrt{\frac{1-4s^2_W}{3}}$\\
$b$ &
$\frac{1}{2}\frac{c_{2W}}{\sqrt{3\left(1-4s^2_W\right)}}$ &
$\frac{1}{2}\frac{1+2s^2_W}{\sqrt{3\left(1-4s^2_W\right)}}$\\ 
$J_1, J_2$ & $\frac{1}{\sqrt{3}}\frac{1-9s^2_W}{\sqrt{1-4s^2_W}}$ & $\frac{1}{\sqrt{3}}\frac{c^2_W}{\sqrt{1-4s^2_W}}$\\
$J_3$ & $-\frac{1}{\sqrt{3}}\frac{\left(1-11s^2_W\right)}{\sqrt{1-4s^2_W}}$ &
$-\frac{1}{\sqrt{3}}\frac{c^2_W}{\sqrt{1-4s^2_W}}$\\ 
\hline 
\end{tabular}
\caption{\label{tab2} The couplings of $Z'$ with fermions.}
\end{table}

\subsection{\label{gsint} Interactions of scalars with gauge bosons}  

Let us note that the interactions of the inert scalars with gauge bosons were given in \cite{dns}. Therefore, in this work we only need to calculate the remaining interactions of the normal scalars with gauge bosons, which are given from $\sum_{S}(D^{\mu}S)^\dagger (D_{\mu}S)$, with $S=\eta,\chi$. Expanding the scalar multiplets in terms of their VEVs and physical scalar fields, $S = \langle S \rangle + S'$, respectively, and noting that they are colorless, i.e. $D_\mu=\pa_\mu+ig P_\mu$ with $P_\mu\equiv P^{\mathrm{CC}}_\mu+P^{\mathrm{NC}}_\mu$, the relevant Lagrangian becomes  
\bea
\sum_{S}(D^{\mu}S)^\dagger (D_{\mu}S) &\supset& \sum_S\left\{\left[ig(\partial^\mu S')^\dagger P_\mu S'+H.c.\right] \right.\crn
&&\left. +\left[g^2\langle S \rangle^\dagger P^\mu P_\mu S' + H.c.\right] + g^2 S'^{\dagger}P^\mu P_\mu S'\right\}.
\eea That being said, the first, second, and last terms in the braces  provide the couplings of two scalars with a gauge boson, two gauge bosons with a scalar, and two gauge bosons with two scalars, respectively. 

We shall work in a basis where all the Goldstone bosons are gauged away. In this unitary gauge, the scalar multiplets simply take the form,
\bea
\eta=\left(\begin{array}{c}\frac{u}{\sqrt{2}}\\0
\\0\end{array} \right) + \left(\begin{array}{c}\frac{c_\xi h+s_\xi H}{\sqrt{2}}\\0\\c_\theta H^+ \end{array}\right),\hs
\chi = \left(\begin{array}{c} 0\\0\\ \frac{w}{\sqrt{2}}\end{array}\right)+\left(\begin{array}{c} s_\theta H^- \\0\\ \frac{-s_\xi h+c_\xi H}{\sqrt{2}}\end{array}\right),\label{physetachi} \eea where in each expansion, the first and second (column) terms correspond to $\langle S \rangle $ and $S'$, respectively. The notations for the scalar multiplets, including the following gauge bosons, in this gauge have conveniently kept unchanged, which should not be confused.    

Considering the first term in the above Lagrangian, 
\be \sum_S\left[ ig(\partial^\mu S')^\dagger P_\mu S'+H.c.\right]=\sum_S\left[  ig\left(\partial^\mu S'\right)^\dagger P^{\mathrm{CC}}_\mu S'+ ig(\partial^\mu S')^\dagger P^{\mathrm{NC}}_\mu S'+H.c.\right],\ee
and substituting $S'$, $P^{\mathrm{CC}}_\mu$, $P^{\mathrm{NC}}_\mu$ as defined above, we get the interactions of a charged or a neutral gauge boson with two normal scalars, as supplied in Table \ref{1g2s}, where throughout the text the interactions are understood as vertex times coupling.   

Continuously, we expand the second term,  
\bea
\sum_S\left[ g^2\langle S \rangle^\dagger P^\mu P_\mu S' + H.c.\right] &=& g^2 \sum_S\left[\langle S\rangle P^{\mathrm{CC}\mu}P_\mu^{\mathrm{CC}} S'+\langle S \rangle \left\{P^{\mathrm{CC}\mu},P_\mu^{\mathrm{NC}}\right\}S'\right.\crn
&&\left. +\langle S \rangle P^{\mathrm{NC}\mu}P^{\mathrm{NC}}_\mu S'+ H.c.\right]. \eea
Substituting the physical fields, the interactions of a scalar with two gauge bosons corresponding to these three terms are resulted as listed in Tables \ref{2cg1s}, \ref{1c1ng1s}, and \ref{2ng1s}, respectively.   
 
For the third term, we have 
\bea 
\sum_S g^2 S'^\dagger P^\mu P_\mu S'=g^2\sum_S\left[S'^\dagger P^{\mathrm{CC}\mu}P^{\mathrm{CC}}_\mu S'+ S'^\dagger\left\{P^{\mathrm{CC}\mu},P_\mu^{\mathrm{NC}}\right\} S' +S'^\dagger P^{\mathrm{NC}\mu}P^{\mathrm{NC}}_\mu S'\right].\eea
These three terms yield the interactions of two scalars with two gauge bosons, as given in Tables \ref{2cg2s}, \ref{1c1ng2s}, and \ref{2ng2s}, respectively.  

\begin{table}[h]
\bc
\begin{tabular}{|c|c|}
\hline 
Vertex & Coupling \\ \hline
$A_{\mu} H^+ \overleftrightarrow{\partial}^\mu H^- $ & $i g s_W$\\
$Z_{\mu} H^- \overleftrightarrow{\partial}^\mu H^+ $ & $ \frac{ig}{2c_W}(s^2_\theta+2s_W^2) $\\
$Z'_{\mu} H^-\overleftrightarrow{\partial}^\mu H^+ $ & $\frac{ig[2(1-4 s_W^2) - s_\theta^2 (1 + 2 s_W^2)]}{2c_W\sqrt{3(1-4s_W^2)}}$\\
$X^+_{\mu} h \overleftrightarrow{\partial}^\mu H^-$ & $ \frac{ig}{2}(c_\theta c_\xi + s_\xi s_\theta )$\\
$X^-_{\mu} H^+ \overleftrightarrow{\partial}^\mu h$ & $\frac{i g}{2}(c_\theta c_\xi + s_\xi s_\theta )$\\
$X^+_{\mu} H^-\overleftrightarrow{\partial}^\mu H$ & $\frac{ig}{2}(s_\theta c_\xi - s_\xi c_\theta)$\\
$X^-_{\mu} H \overleftrightarrow{\partial}^\mu H^+$ &$\frac{i g}{2}(s_\theta c_\xi - s_\xi c_\theta)$\\
\hline
\end{tabular}
\caption{\label{1g2s}
The interaction of a gauge boson with two normal scalars.}
\ec
\end{table}

\begin{table}[h]
\bc
\begin{tabular}{|c|c|}
\hline 
Vertex & Coupling  \\ \hline
$hW^+_{\mu}W^{\mu-}$ & $ \frac{g^2}{2}uc_\xi $ \\
 $hX^+_{\mu}X^{\mu-}$  & $\frac{g^2}{2}(uc_\xi - ws_\xi )$\\
$hY^{++}_{\mu}Y^{\mu--}$ & $\frac{-g^2}{2}ws_\xi$\\
$HW^+_{\mu}W^{\mu-} $ &$\frac{g^2}{2}us_\xi$\\
$HX_{\mu}^-X^{\mu+} $ & $ \frac{g^2}{2}(us_\xi+wc_\xi) $\\
$HY^{++}_{\mu}Y^{\mu--}$ & $\frac{g^2}{2}wc_\xi$\\
$H^+W^+_{\mu}Y^{\mu--} $ &$\frac{g^2}{2\sqrt{2}}(uc_\theta+ws_\theta)$\\
$H^-W^-_{\mu}Y^{\mu++}$ & $\frac{g^2}{2\sqrt{2}}(uc_\theta+ws_\theta)$\\
\hline
\end{tabular}
\caption{\label{2cg1s}
The interaction of two charged gauge bosons with a normal scalar.}
\ec
\end{table}

\begin{table}[h]
\bc
\begin{tabular}{|c|c|}
\hline 
Vertex & Coupling \\ \hline
$A_{\mu}X^{\mu-}H^+$ & $ \frac{g^2}{2}s_W(uc_\theta-ws_\theta )$ \\ 
$A_{\mu}X^{\mu+}H^-$  & $\frac{g^2}{2}s_W(uc_\theta-ws_\theta )$\\
$Z_{\mu}X^{\mu-}H^+$  & $\frac{g^2}{4c_W}\left[(1-2s_W^2 )uc_\theta+ws_\theta(1+2s_W^2)\right]$\\
$Z_{\mu}X^{\mu+}H^-$  & $\frac{g^2}{4c_W}\left[(1-2s_W^2)uc_\theta+ws_\theta(1+2s_W^2)\right]$\\
$Z'_{\mu}X^{\mu+}H^-$  & $\frac{g^2}{4c_W\sqrt{3(1-4s_W^2)}}\left[(-1+4s_W^2 )uc_\theta-ws_\theta(1+8s_W^2)\right]$\\
$Z'_{\mu}X^{\mu-}H^+$  & $\frac{g^2}{4c_W\sqrt{3(1-4s_W^2)}}\left[(-1+4s_W^2 )uc_\theta-ws_\theta(1+8s_W^2)\right]$\\
\hline
\end{tabular}
\caption{\label{1c1ng1s}
The interaction of a charged and a neutral gauge boson with a normal scalar.}
\ec
\end{table}

\begin{table}[h]
\bc
\begin{tabular}{|c|c|}
\hline 
Vertex & Coupling \\ \hline
$hZ_{\mu}Z^{\mu}$  & $\frac{g^2}{4c_W^2}uc_\xi$\\
$hZ_{\mu}Z'^{\mu}$  & $\frac{g^2}{2\sqrt{3}c_W^2}\sqrt{1-4s_W^2}uc_\xi$\\
$hZ'_{\mu}Z'^{\mu}$  & $\frac{g^2}{12c_W^2(1-4s_W^2)}\left[(1-4s_W^2)^2uc_\xi-4ws_\xi c_W^4\right]$\\
$HZ_{\mu}Z^{\mu}$ & $ \dfrac{g^2}{4c_W^2}us_\xi$ \\
$HZ_{\mu}Z'^{\mu}$  & $\frac{g^2}{2\sqrt{3}c_W^2}\sqrt{1-4s_W^2}us_\xi$\\
 $HZ'_{\mu}Z'^{\mu}$  & $\frac{g^2}{12c_W^2(1-4s_W^2)}\left[u(1-4s_W^2)^2s_\xi+4wc_W^4c_\xi \right]$\\
\hline
\end{tabular}
\caption{\label{2ng1s}
The interaction of two neutral gauge bosons with a normal scalar.}
\ec
\end{table}

\begin{table}[h]
\bc
\begin{tabular}{|c|c|c|c|}
\hline 
Vertex & Coupling & Vertex & Coupling \\ \hline
$W^+_{\mu}W^{\mu-}hh $ & $\frac{g^2}{4}c^2_\xi$ & $X^+_{\mu}X^{\mu-}hh$ & $\frac{ g^2}{4} $  \\
$W^{+}_{\mu}W^{\mu-}Hh$ &  $\frac{g^2}{2}s_\xi c_\xi$  &   $X^{+}_{\mu}X^{\mu-}HH$ & $\frac{g^2}{4}$ \\
$W^{+}_{\mu}W^{\mu-}HH$ & $\frac{g^2}{4}s^2_\xi$   &  $ X^{+}_{\mu}X^{\mu-}H^{+}H^{-}$ & $\frac{g^2}{2}$  \\
 $W^{+}_{\mu}W^{\mu-} H^{+}H^{-} $ & $\frac{g^2}{2}s_\theta^2$     &  $Y^{++}_{\mu}Y^{\mu--}hh$ & $\frac{ g^2}{4}s^2_\xi$\\
$Y^{++}_{\mu}W^{\mu-}H^-H$ & $\frac{g^2}{2\sqrt{2}} s_{\theta + \xi}$ & $Y^{++}_{\mu}Y^{\mu--}Hh$ & $-\frac{ g^2}{2}c_\xi s_\xi$  \\
$ Y^{++}_{\mu}W^{\mu-}H^-h$ & $\frac{g^2}{2\sqrt{2}}c_{\theta +\xi}$ & $ Y^{++}_{\mu}Y^{\mu--}HH $ & $ \frac{g^2}{4}c^2_\xi $  \\
$XXHh$ & 0 & $Y^{++}_{\mu}Y^{\mu--}H^+H^-$ & $\frac{g^2}{2}c_\theta^2 $ \\
\hline
\end{tabular}
\caption{\label{2cg2s}
The interaction of two charged gauge bosons with two scalars.}
\ec
\end{table}

\begin{table}[h]
\bc
\begin{tabular}{|c|c|}
\hline 
Vertex & Coupling \\ \hline
$A_{\mu}X^{\mu-}H^+h$ & $ \frac{g^2}{2}s_W c_{\xi-\theta}$ \\
$A_{\mu}X^{\mu-}H^+H$ & $ \frac{g^2}{2}s_W s_{\xi - \theta}$ \\
$Z_{\mu}X^{\mu-}H^+h$  & $\frac{g^2}{4c_W}[c_\theta c_\xi \left(1-2s_W^2\right)-s_\theta s_\xi\left(1+2s_W^2\right)]$\\
$Z_{\mu}X^{\mu-}H^+H$ & $ \frac{g^2}{4c_W}\left[\left(1+2s_W^2\right)c_\xi s_\theta + \left(1-2s_W^2\right)s_\xi c_\theta\right]$ \\
$Z'_{\mu}X^{\mu-}H^+h$  & $-\frac{g^2}{4c_W\sqrt{3\left(1-4s^2_W\right)}}\left[\left(1-4s_W^2\right)c_\theta c_\xi -s_\theta s_\xi\left(1+8s_W^2\right)\right]$\\
$Z'_{\mu}X^{\mu-}H^+H$  & $-\frac{g^2}{4c_W\sqrt{3\left(1-4s^2_W\right)}}\left[ \left(1+8s_W^2\right)c_\xi s_\theta+c_\theta s_\xi\left(1-4s_W^2\right)\right]$\\
\hline
\end{tabular}
\caption{\label{1c1ng2s}
The interactions of two gauge bosons with two scalars.}
\ec
\end{table}

\begin{table}[h]
\bc
\begin{tabular}{|c|c|}
\hline 
Vertex & Coupling \\ \hline
$Z_{\mu} Z^{\mu} hh$&$\frac{ g^2}{8 c_W^2}c_\xi^2$\\
$Z_{\mu} Z^{\mu} Hh$ & $\frac{g^2 }{4 c_W^2}c_\xi  s_\xi$\\
$Z_{\mu} Z^{\mu} HH$ & $\frac{g^2 }{8 c_W^2}s_\xi^2$\\
$Z_{\mu} Z'^{\mu} hh$&$\frac{g^2}{4\sqrt{3}c_W^2}\sqrt{1-4s_W^2}c_\xi^2$\\
$Z_{\mu} Z'^{\mu}HH$ & $\frac{g^2}{4\sqrt{3}c_W^2}\sqrt{1-4s_W^2}s_\xi^2$\\
$Z'_{\mu} Z'^{\mu} hh$&$\frac{g^2}{24 c_W^2 \left(1-4 s_W^2\right)} \left[\left(1-4s_W^2\right)^2c_\xi^2+4s_\xi^2c_W^4\right]$\\
$Z'_{\mu} Z'^{\mu} HH$ & $\frac{g^2}{24 c_W^2 \left(1-4 s_W^2\right)} \left[\left(9-24c_W^2+16c_W^4\right)s_\xi^2+4c^2_\xi c_W^4\right]$\\
$Z'_{\mu} Z'^{\mu} Hh$&$\frac{g^2}{4 c_W^2 (1-4 s_W^2)}\left(3 - 8 c_W^2 + 4 c_W^4\right)s_\xi c_\xi$\\
$A_{\mu} A^{\mu} H^+H^-$ & $ g^2 s_W^2$\\
$A_{\mu} Z^{\mu} H^+H^-$ & $-\frac{ g^2s_W}{c_W} \left[2 s_W^2 c_\theta^2+ s_\theta^2\left(1+2s^2_\theta\right)\right]$\\
$Z_{\mu} Z^{\mu} H^+H^-$ & $ \frac{g^2}{4 c_W^2} \left[4s_W^4c_\theta^2+s_\theta^2\left(1+2s_W^2\right)^2\right]$\\
$A_{\mu} Z'^{\mu} H^+H^-$ & $-\frac{g^2s_W}{c_W \sqrt{3-12 s_W^2}} \left[\left(1-10s^2_W\right)s_\theta^2+2 c_\theta^2\left(1-4s^2_W\right)\right]$\\ 
$Z_{\mu} Z'^{\mu} H^+H^-$ & $-\frac{g^2}{2 c_W^2 \sqrt{12 c_W^2-9}}\left[4 \left(3 - 7 c_W^2 + 4 c_W^4\right) c^2_\theta+s^2_\theta\left(27 - 48 c_W^2 + 20 c_W^4\right)\right]$\\
$Z'_{\mu} Z'^{\mu} H^+H^-$ & $\frac{g^2}{12c^2_W\left(1-4s_W^2\right)} \left[ 4 \left(1-4s^2_W\right)^2 c^2_\theta+s^2_\theta\left(1-10s_W^2\right)^2\right]$\\
\hline
\end{tabular}
\caption{\label{2ng2s}
The interaction of two neutral gauge bosons with two scalars.}
\ec
\end{table}

\subsection{Scalar self-interactions and Yukawa interactions}

Since we work in the unitary gauge, the scalar self-interactions include only those with physical scalar particles. Note that the interactions between the normal scalars and the inert scalars were given in \cite{dns}. Therefore, we necessarily calculate only self-interactions of the normal scalars. That being said, substituting $\eta$ and $\chi$ from (\ref{physetachi}) into $V_{\mathrm{simple}}$, we obtain the relevant interactions as given in Tables \ref{sio3s} and \ref{sio4s}. 

\begin{table}[h]
\bc
\begin{tabular}{|c|c|}
\hline 
Vertex & Coupling \\ \hline
$hhh$ & $ -\frac{1}{2}\left[u c_\xi\left(2\lambda_1 c_\xi^2+s_\xi^2\lambda_3\right)-w s_\xi \left(\lambda_3c_\xi^2  + 2s_\xi^2\lambda_2\right)  \right]$ \\
$HHH$ & $ -\frac{1}{2}\left[u s_\xi\left(2\lambda_1 s_\xi^2+c_\xi^2\lambda_3\right) + w c_\xi \left(\lambda_3s_\xi^2  + 2c_\xi^2\lambda_2\right)  \right] $\\
$hhH$ & $-\frac{1}{2}\left\{\left[\lambda_3 s_\xi^2+2c_\xi^2\left(3\lambda_1-\lambda_3\right)\right] u s_\xi+ w c_\xi \left[\lambda_3c_\xi^2+2s_\xi^2 \left(3\lambda_2-\lambda_3\right)\right]\right\}$\\
$hHH$ & $-\frac{1}{2}\left\{ \left[\lambda_3 c_\xi^2+2s_\xi^2\left(3\lambda_1-\lambda_3\right)\right]u c_\xi- w s_\xi \left[\lambda_3s_\xi^2+2c_\xi^2 \left(3\lambda_2-\lambda_3\right)\right]\right\}$\\
$hH^+H^-$ & $\frac{1}{2}\lambda_4s_{2\theta}\left(s_\xi u - c_\xi w\right)+s_\theta^2 \left[2 w\lambda_2 s_\xi -u c_\xi 
\left(\lambda_3 + \lambda_4\right)\right] + 
 c_\theta^2 \left[ w\left(\lambda_3 + \lambda_4\right) s_\xi -2u c_\xi \lambda_1\right]$ \\
  $HH^+H^-$ & $ - \frac{1}{2}\lambda_4 s_{2\theta}\left (c_\xi u + s_\xi w\right) - s_\theta^2 \left[2w\lambda_2c_\xi + u s_\xi\left(\lambda_3  + \lambda_4\right) \right] - c_\theta^2 \left[w(\lambda_3 + \lambda_4)c_\xi+ 2u s_\xi\lambda_1  \right]$\\
\hline
\end{tabular}
\caption{\label{sio3s}
The self-interaction of three normal scalars.}
\ec
\end{table}

\begin{table}[h]
\bc
\begin{tabular}{|c|c|}
\hline 
Vertex & Coupling \\ \hline
$hhhh$ & $-\frac{1}{16}\left[4\left(\lambda_2s_\xi^4+\lambda_1c_\xi^4\right)+\lambda_3s_{2\xi}^2\right] $\\ 
$HHHH$ & $ -\frac{1}{16}\left[4\left(\lambda_2c_\xi^4+\lambda_1s_\xi^4\right)+\lambda_3s_{2\xi}^2\right] $ \\
$H^+H^-H^+H^-$ & $-\left(\frac{1}{4}\lambda_3 s_{2\theta}^2+\lambda_1c_\theta^4+\lambda_2s_\theta^4\right)$ \\
$hhHH$ & $-\frac{1}{8}\left[2\lambda_3+3s_{2\xi}^2\left(\lambda_1+\lambda_2-\lambda_3\right)  \right] $ \\
$hhhH$ & $-\frac{1}{4}s_{2\xi}\left[\left(\lambda_3-2\lambda_2\right)s_\xi^2+c_\xi^2\left(2\lambda_1-\lambda_3\right)\right] $ \\
$hHHH$&$-\frac{1}{4}s_{2\xi}\left[\left(\lambda_3-2\lambda_2\right)c_\xi^2+s_\xi^2\left(2\lambda_1-\lambda_3\right)\right]$\\
$hHH^+H^-$ & $ -\frac{1}{2}\left[\left(2\lambda_1-\lambda_3-\lambda_4\right)s_{2\xi}c_\theta^2+s_\theta^2 s_{2\xi}\left(\lambda_3+\lambda_4-2\lambda_2\right)+\lambda_4s_{2\theta}c_{2\xi}\right] $\\
$hhH^+H^-$ & $ - \frac{1}{4}\left\{-\lambda_4s_{2\theta}s_{2\xi}+2c_\theta^2\left[\left(\lambda_3+\lambda_4\right)s^2_\xi  +2\lambda_1c^2_\xi\right]+2s_\theta^2\left[\left(\lambda_3+\lambda_4\right) c_\xi^2+2\lambda_2s_\xi^2\right]\right\}$\\
$HHH^+H^-$ & $ - \frac{1}{4}\left\{\lambda_4s_{2\theta}s_{2\xi}+2c_\theta^2\left[\left(\lambda_3+\lambda_4\right) c_\xi^2+2\lambda_1s_\xi^2\right]+2s_\theta^2\left[\left(\lambda_3+\lambda_4\right)s^2_\xi +2\lambda_2c^2_\xi \right]\right\}$\\
\hline
\end{tabular}
\caption{\label{sio4s}
The self-interaction of four normal scalars.}
\ec
\end{table}

The inert scalars do not have Yukawa interaction with fermions due to $Z_2$ symmetry. Therefore, we turn to investigate the Yukawa interactions of the normal scalars. For this aim, we first identify the mass matrices of fermions:
\bea  
 && m_{3b}^u = -\frac{h_{3b}^u}{\sqrt{2}}u,\hs  m_{\al b}^u = -\frac{h_{\alpha b}^u}{2\Lambda}uw,\\
&& m_{3b}^d = \frac{h_{3b}^d}{2\Lambda}uw,\hs  m_{\al b}^d= -\frac{h_{\alpha b}^d}{\sqrt{2}}u,\\
&& m_{33}^J = -\frac{h_{33}^J}{\sqrt{2}}w,\hs  m_{\alpha \beta}^J= -\frac{h_{\alpha \beta}^J}{\sqrt{2}}w,\label{exmass} \\
&& m^e_{ab}=\sqrt{2}u\left(h^e_{ab}+\frac{h'^{e}_{ba}w^2}{4\Lambda^2}\right), 
\eea where the irrelevant neutrino masses, which have not been listed, can be founded in \cite{dns}. 
Hence, the relevant interactions and couplings are resulted as in Tables \ref{hinteractions}, \ref{Hinteractions}, and \ref{Hcinteractions}. When the above mass matrices are diagonalized, we have such similar interactions for the physical fields where the Yukawa couplings depend only on the mass eigenvalues.      

\begin{table}[h]
\bc
\begin{tabular}{|c|c|}
\hline 
Vertex & Coupling \\ \hline
$h\bar{e}_ae_b$  & $-\frac{m^e_{ab}}{u}c_\xi + \frac{h'^{e}_{ba}}{\sqrt{2}\Lambda^2}w u s_\xi$\\
$h\bar{u}_3u_3$  & $-\frac{m_{33}^u}{u} c_\xi $ \\
$h\bar{u}_\alpha u_\alpha$  & $ -m_{\alpha\alpha}^u (\frac{c_\xi}{u} -\frac
{s_\xi}{w})$ \\
$h\bar{d}_3d_3$  & $-m_{33}^d (\frac{c_\xi}{u} - \frac{s_\xi}{w} )$ \\
$h\bar{d}_\alpha d_\alpha$  & $-\frac{m_{\alpha\alpha}^d}{u} c_\xi$ \\
$h\bar{J}_3J_3$&$\frac{m_{33}^J}{w}s_\xi$\\
$h\bar{J}_{\alpha} J_{\beta}$&$\frac{m_{\alpha\beta}^J}{w}s_\xi$\\
\hline
\end{tabular}
\caption{\label{hinteractions}
The Yukawa interactions of the standard model Higgs like particle ($h$).}
\ec
\end{table}

\begin{table}[h]
\bc
\begin{tabular}{|c|c|}
\hline 
Vertex & Coupling \\ \hline
$H\bar{e}_a e_b$  & $-\frac{m^e_{ab}}{u} s_\xi - \frac{h'^{e}_{ba}}{\sqrt{2}\Lambda^2}w u c_\xi$\\
$H\bar{u}_3 u_3$  & $-\frac{m_{33}^u}{u} s_\xi $ \\
$H\bar{u}_\alpha u_\alpha$  & $-m_{\alpha\alpha}^u (\frac{c_\xi}{w} + \frac{s_\xi}{u} )$ \\
$H\bar{d}_3 d_3$  & $-m_{33}^d (\frac{c_\xi}{w} + \frac{s_\xi}{u} )$ \\
$H\bar{d}_\alpha d_\alpha$  & $ -\frac{m_{\alpha\alpha}^d}{u}s_\xi$ \\
$H\bar{J_3}J_3$&$-\frac{m_{33}^J}{w}c_\xi$\\
$H\bar{J}_{\alpha}J_\beta$&$-\frac{m_{\alpha\beta}^J}{w}c_\xi$ \\
\hline
\end{tabular}
\caption{\label{Hinteractions}
The Yukawa interactions of the new neutral Higgs boson ($H$).}
\ec
\end{table}

\begin{table}[h]
\bc
\begin{tabular}{|c|c|}
\hline 
Vertex & Coupling \\ \hline
$H^+\bar{J_3}u_a$ & $-\frac{m_{3a}^u}{u}\sqrt{2}c_\theta$\\
$H^-\bar{J_\alpha}d_a$&$-\frac{m_{\alpha a}^d}{ u}\sqrt{2}c_\theta$\\
$H^-\bar{u}_3J_3$&$-\frac{m_{33}^J}{w}\sqrt{2}s_\theta$\\
$H^+\bar{d_\alpha}J_\beta$&$-\frac{m_{\alpha \beta}^J}{w}\sqrt{2}s_\theta$\\
$H^+H^-\bar{u}_\alpha u_a$&$\frac{m_{\alpha a}^u}{ uw}2s_\theta c_\theta $\\
$H^+H^-\bar{d_3}d_a$&$\frac{m_{3a}^d}{uw}2s_\theta c_\theta $\\
\hline
\end{tabular}
\caption{\label{Hcinteractions}
The Yukawa interactions of the charged Higgs boson ($H^\pm$).}
\ec
\end{table}

\section{\label{pheno} Phenomenology}

\subsection{\label{higgs} The standard model like Higgs particle}

The discovery of the Higgs particle marks the success of the LHC run I \cite{higgsexp}, and its couplings can be summarized via the combined best-fit signal strength, $\mu_h=1.1\pm0.1$, which deviates 10\% from the standard model value of 1 \cite{newdatahiggs}\footnote{See also \cite{uhaischhiggs} for an intriguing discussion.}. Let us particularly investigate the Higgs coupling to two photons that substitutes in 
\be \mu_{\ga\ga}=\fr{\sigma(pp\rightarrow h) Br(h\rightarrow \ga\ga)}{\sigma(pp\rightarrow h)_{\mathrm{SM}} Br(h\rightarrow \ga\ga)_{\mathrm{SM}}}, \ee where the numerator is given by the considering model once measured by the experiments, while the denominator is the standard model prediction. Since the LHC run II data from ATLAS and CMS \cite{run2cbh} yield $\mu_{\gamma\gamma}$ mostly coinciding with the run I combined signal, the $\mu_h$ strength can be taken as the benchmark value for further investigation on this channel. 

\begin{figure}[h]
\begin{center}
\includegraphics{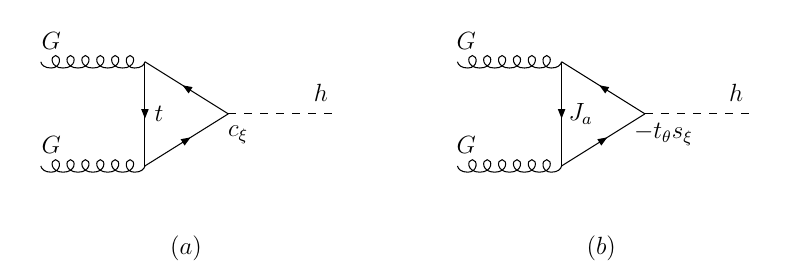}
\caption[]{\label{ggf} Contributions to the Higgs production due to gluon-gluon fusion. Here, the $h$-$H$ mixing effect changes the $h\bar{t}t$ coupling by $c_\xi$ and simultaneously couples $h$ to the exotic quarks which run in the loop. Such two modifications of the new physics are comparable $\sim (u/w)^2$. The $h\bar{f}f$ coupling for $f=t,J_a$ is normalized to the standard model coupling, i.e. $h_f=-\fr{m_f}{u}g_f$, in which $g_f$ is indicated in the relevant graph.}  
\end{center}
\end{figure}

The Higgs production dominantly comes from the gluon gluon fusion via top loops \cite{higgsprodecay}. Hence, we can approximate  $\sigma(pp\rightarrow h)\simeq \sigma(GG \rightarrow h)$ as given in Fig \ref{ggf}, where the new physics effects are included. Note that the (b) diagram was skipped in \cite{RM331Queiroz,diphotonHiggs}. We have
\be \fr{\sigma(pp\rightarrow h)}{\sigma(pp\rightarrow h)_{\mathrm{SM}}}\simeq \left|\fr{c_\xi A_t(\tau_t)- t_\theta s_\xi \sum_J A_J (\tau_J)}{A_t(\tau_t)}\right|^2,\ee where $A_f(\tau_f)=2\left[\tau_f+(\tau_f-1)\arcsin^2(\sqrt{\tau_f})\right]/\tau^2_f$ and $\tau_f=m^2_h/(4m^2_f)<1$ for $f=t,J_{1,2,3}$.             

\begin{figure}[h]
\begin{center}
\includegraphics{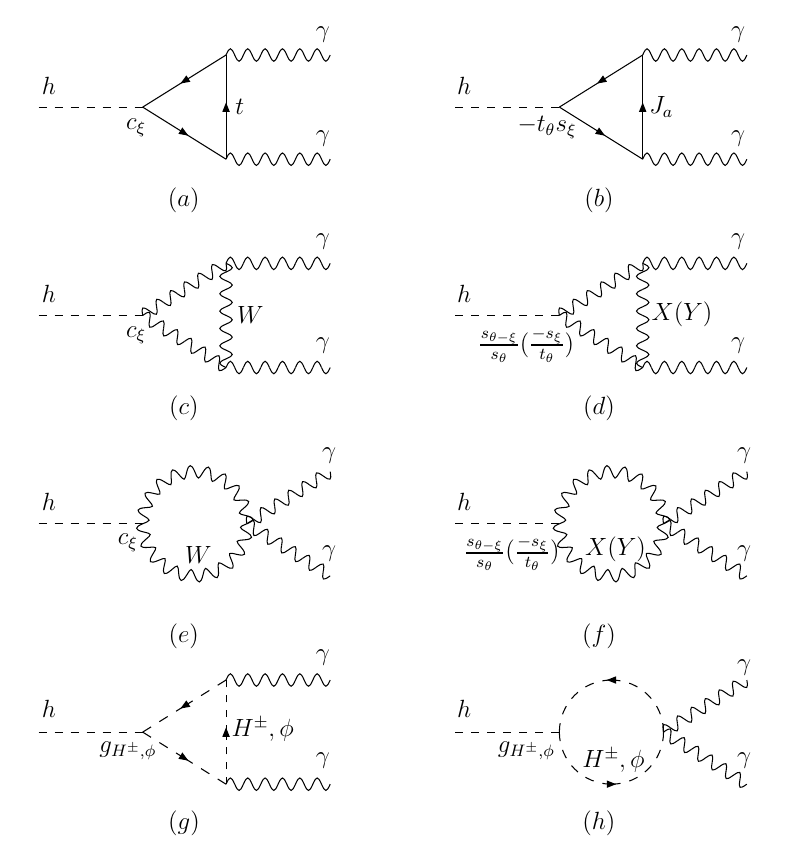}
\caption[]{\label{hgg} Contributions to the decay $h\rightarrow \ga\ga$. Here, the new physics effects include the $h$-$H$ mixing that modifies the relevant couplings and the contributions of the exotic quarks, new charged gauge bosons, and charged normal and inert scalars. Hereafter, the couplings $hV^*V$ $(V=W,X,Y)$ and $hS^*S$ ($S=H^\pm,\phi$) are normalized to the standard model couplings, i.e. $h_V=\fr{g^2u}{2} g_V$ and $h_S=-\fr{g^2u}{2} g_S$, in which $g_V$ is explicitly displayed in the relevant graph whereas $g_{H^\pm}$ and $g_\phi$ could be read off from Table \ref{sio3s} and those in~\cite{dns}, respectively.}   
\end{center}
\end{figure}

The dominant contributions and the new physics effects to the Higgs decay into two photons are given in Fig. \ref{hgg}. The decay width is  
\be \Ga(h\rightarrow \ga\ga)=\fr{G_F \al^2 m^3_h}{128\sqrt{2}\pi^3}\left|\sum_f N_C Q^2_f g_f A_f+\sum_{B=V,S} \fr{m^2_W}{m^2_B}Q^2_B g_B A_B \right|^2, \ee where $f=(t,J_{1},J_2,J_3)$, $V=(W, X, Y)$, $S=(H,\phi)$, $\sqrt{2}G_F=1/u^2$, and $\al=e^2/(4\pi)$. $N_C$ and $Q_{f,V,S}$ correspond to the color factor and electric charges of the fields, respectively. $A_f$ is given as before, while $A_{V,S}$ take the respective forms \cite{higgsprodecay},
\bea
&& A_V(\tau_V)=-\left[2 \tau^2_V+3\tau_V+3(2\tau_V-1)f(\tau_V)\right]/\tau_V^2,\crn
&& A_S(\tau_S)=-\left[\tau_S-f(\tau_S)\right]/\tau_S^2, 
\eea where all the new particles have natural masses beyond the weak scale \cite{dns}, thus $\tau_B=m^2_h/(4m^2_B)<1$ and $f(\tau_B)=\arcsin^2(\sqrt{\tau_B})$. 

The total Higgs decay width is significantly summed over channels,
\be \Ga(h\rightarrow \mathrm{all})=\sum_{f=b,c,\tau} \Ga(h\rightarrow \bar{f}f)+\sum_{V=W,Z}\Ga(h\rightarrow VV^*)+\Ga(h\rightarrow \ga\ga)+\Ga(h\rightarrow GG). \ee With the aid of \cite{higgsprodecay}, we obtain
\ben \item Fermion modes:  
\be \Ga(h\rightarrow \bar{f}f)=\fr{G_F N_C}{4\sqrt{2}\pi}m_h g^2_f m^2_f\left(1-1/\tau_f\right)^{3/2},\ee where $g_f=c_{\xi+\theta}/c_\theta$ for $f=b,c$ and $g_f=c_\xi-(1/\sqrt{2})(w/\La)^2(h'_\tau u/m_\tau)s_\xi t_\theta$ for $f=\tau$, where $h'_\tau$ is the 33 component of $h'^e_{ab}$ in the mass eigenstates.   
\item Weak-boson modes:
\be \Ga(h\rightarrow VV^*)=\fr{3G^2_F m^4_V m_h}{16\pi^3}c^2_\xi \delta_V R(x),\ee where $\delta_W=1$, $\delta_Z=7/12-10s^2_W/9+40s^4_W/27$, $x=m^2_V/m^2_h$, and
\be R(x)=\fr{3(1-8x+20x^2)}{\sqrt{4x-1}}\arccos\left(\fr{3x-1}{2x\sqrt{x}}\right)-\fr{1-x}{2x}(2-13x+47x^2)-\fr 3 2 (1-6x+4x^2)\ln x.  \ee
\item Gluon mode:
\be \Ga(h\rightarrow GG)=\fr{G_F\al^2_s m^3_h}{36\sqrt{2}\pi^3}\left|\fr 3 4 \left[c_\xi A_t (\tau_t)-t_\theta s_\xi \sum_J A_J(\tau_J)\right] \right|^2,\ee where the form factors are defined above.    
\een

The corresponding standard model Higgs decay widths are those obtained above in the limit $w\rightarrow \infty,\ \theta\rightarrow 0,\ \xi \rightarrow 0$ such that $\Ga(\cdots)_{\mathrm{SM}}=\Ga(\cdots)|_{\theta=\xi=0,w=\infty}$. The fixed input parameters for estimating the signal strength are $u\simeq 246$ GeV, $s^2_W\simeq 0.231$, $\al\simeq 1/128$, $\al_s\simeq 0.118$, $m_{\tau}\simeq 1.776$ GeV, $m_b\simeq 4.18$ GeV, $m_c\simeq 1.275$ GeV, $m_t\simeq 173$ GeV, $m_h\simeq 125\ \mathrm{GeV}$, $\La=5$ TeV, and $w=3.6$--5 TeV, which should satisfy both the FCNC bound and the Landau pole limit. Furthermore, we have the conditions for potential parameters, 
\bea && u^2=\fr{2(2\la_2 \mu_1^2-\la_3\mu_2^2)}{\la^2_3-4\la_1\la_2},\hs w^2=\fr{2(2\la_1 \mu_2^2-\la_3\mu_1^2)}{\la^2_3-4\la_1\la_2},\\ 
&& \mu^2_{1,2}<0,\hs \la_{1,2,4}>0,\hs -2\sqrt{\la_1\la_2}<\la_3<\mathrm{Min}\left\{2\la_1\left({\mu_2}/{\mu_1}\right)^2,2\la_2\left({\mu_1}/{\mu_2}\right)^2\right\},\eea and the others for the inert part as supplied in \cite{dns}. 

The signal strength $\mu_{\ga\ga}$ is scanned for $w=3.6$, 4, 4.5, 5 TeV when $\la_{1,2,4}$ ($\la_3$ is related to $\la_{1,2}$ by $m_h$) vary in 0.01--2 but satisfy all the above conditions, while the coupling and mass parameters of the inert part are chosen as in \cite{dns}. The exotic quark masses are also varied from a value above the weak scale up to the TeV scale. Our numerical study yields a maximal bound, $1 \leq \mu_{\gamma\gamma}\leq 1.06$, in agreement with the data (up to the known QCD corrections). It is clear that the new physics is quite decoupled because the mixing angles $\theta\simeq u/w$ and $\xi\simeq (\la_3 u)/(2\la_2 w)$ are strictly suppressed, while all the contributions of new particles to the amplitudes are proportional to $(u/w)^2$ which are prevented too.

\subsection{\label{Bsdecay} The $B_s$-$\bar{B}_s$ mixing and rare $B_s\rightarrow \mu^+\mu^-$ decay}

Let the gauge states for up-quarks as $u=(u_1\ u_2\ u_3)^T$ and down-quarks as $d=(d_1\ d_2\ d_3)^T$. They are related to the mass eigenstates, $u_{L,R}=V_{uL,R}u'_{L,R}$ and $d_{L,R}=V_{dL,R}d'_{L,R}$, such that $V^\dagger_{uL} m^u V_{uR}=\mathrm{diag}(m_u,m_c,m_t)$ and $V^\dagger_{dL} m^d V_{dR}=\mathrm{diag}(m_d,m_s,m_b)$, where we denote $u'=(u\ c\ t)^T$ and $d'=(d\ s\ b)^T$. The CKM matrix is $V_{\mathrm{CKM}}=V^\dagger_{uL}V_{dL}$. Below, we also denote $q$ to be either $u$ or $d$, while $q'$ is either $u'$ or $d'$. 

Since the quark generations are nonuniversal under the $SU(3)_L\otimes U(1)_X$ gauge symmetry, there must be the corresponding tree-level FCNCs. Indeed, reconsidering the interaction of neutral gauge bosons with fermions (\ref{fin1}), we have 
\bea \mathcal{L}_{\mathrm{NC}}&=& -g\bar{F}\ga^\mu P^{\mathrm{NC}}_\mu F\crn
&=&-g\bar{F}\ga^\mu[T_3 A_{3\mu}+T_8 A_{8\mu}+t (Q-T_3+\sqrt{3}T_8) B_\mu]F\crn
&\supset& -g\bar{F}\ga^\mu T_8 F (A_{8\mu}+\sqrt{3} t B_\mu),\eea where $t=g_X/g$, $Q-T_3+\sqrt{3}T_8=X$, and $F$ runs over all the fermion multiplets (the sum notation was omitted and should be understood). Above, note that there is no FCNC associated with $T_3$ and $Q$ since all the repetitive flavors, e.g. $\{ u_{aL}\}$, $\{u_{aR}\}$, or $\{D_{\al L}\}$, are identical under these charges. Further, the repetitive flavors of leptons and exotic quarks are also identical under $T_8$. Hence, there are only FCNCs associated with $T_8$ for ordinary quarks. That being said, the relevant interactions are 
\be \mathcal{L}_{\mathrm{NC}} \supset -\fr{g}{\sqrt{1-3t^2_W}}\bar{q}_{aL} \ga^\mu T_8(q_{aL}) q_{aL} Z'_\mu=-\fr{g}{\sqrt{1-3t^2_W}}\bar{q}_{L} \ga^\mu T_{q} q_{L} Z'_\mu, \ee where we have used $A_8+\sqrt{3}t B=Z'/\sqrt{1-3t^2_W}$, $T_8(q_{aR})=0$, and $T_{q}=\fr{1}{2\sqrt{3}}\mathrm{diag}(-1,-1,1)$ that consists of the $T_8$ values for either $(u_{1L}, u_{2L},u_{3L})$ or $(d_{1L},d_{2L},d_{3L})$ flavors, respectively.               
Changing to the mass eigenstates yields the tree-level FCNCs,  
\bea \mathcal{L}_{\mathrm{FCNC}} &=& -\fr{g}{\sqrt{1-3t^2_W}}\bar{q}'_{iL} \ga^\mu (V^\dagger_{qL}T_{q} V_{qL})_{ij} q'_{jL} Z'_\mu\crn
&=&-\fr{g}{\sqrt{3}\sqrt{1-3t^2_W}}[(V^*_{qL})_{3i}(V_{qL})_{3j}]\bar{q}'_{iL}\ga^\mu q'_{jL} Z'_\mu\hs (i\neq j).\label{fin2} \eea

\begin{figure}[h]
\begin{center}
\includegraphics{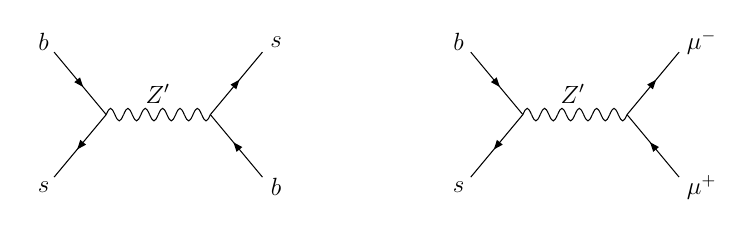} 
\caption[]{\label{dsmm} Contributions to the $B_s$-$\bar{B}_s$ mixing and rare $B_s \rightarrow \mu^+\mu^-$ decay due to the tree-level flavor-changing coupling, $Z'\bar{s}b$, as a character for this kind of the models.}
\end{center}
\end{figure}

Integrating $Z'$ out from (\ref{fin2}), we obtain the effective interactions,
\be \mathcal{H}_{\mathrm{FCNC}}^{\mathrm{eff (a)}}\simeq \fr{[(V^*_{qL})_{3i}(V_{qL})_{3j}]^2}{w^2}(\bar{q}'_{iL}\ga^\mu q'_{jL})^2,\label{fin3}\ee
where we have used $m^2_{Z'}\simeq \fr{g^2w^2}{3(1-3t^2_W)}$. It is noteworthy that the interactions of $Z'$ in (\ref{fin1}) and (\ref{fin2}) may encounter a Landau pole, $\La\sim 5$ TeV, at which $s^2_W(\La)=1/4$ or $g_X(\La)=gs_W/\sqrt{1-4s^2_W}=\infty$ \cite{landau}. However, the effective interactions (\ref{fin3}) are always independent of this singularity. Such interactions contribute to neutral meson mixings as $K$-$\bar{K}$, $D$-$\bar{D}$, $B_d$-$\bar{B}_d$, and $B_s$-$\bar{B}_s$, governed by quark pairs $(q'_i,q'_j)=(d,s),\ (u,c),\ (d,b)$, and $(s,b)$, respectively. The strongest bound for the new physics comes from $B_s$-$\bar{B}_s$ mixing. See the left graph of Fig. \ref{dsmm} for this mixing as explained by basic $Z'$ boson. We have \cite{pdg}
\be \fr{[(V^*_{dL})_{32} (V_{dL})_{33}]^2}{w^2} < \fr{1}{(100\ \mathrm{TeV})^2}.\ee Without loss of generality, let $u_a$ be flavor-diagonal, i.e. $V_{\mathrm{CKM}}=V_{dL}$. The CKM factor is given by $|(V^*_{dL})_{32} (V_{dL})_{33}|\simeq 3.9\times 10^{-2}$ \cite{pdg}, which implies \be w>3.9\ \mathrm{TeV},\ee slightly larger than the bound given in \cite{dns}. 

Correspondingly, the $Z'$ mass is bounded by $m_{Z'}>4.67$ TeV, provided that $s^2_W\simeq 0.231$ is at the low energy regime of the interested precesses. This mass is close to the energy point at which the perturbative character of the $U(1)_X$ (thus $Z'$) interaction is lost. Beyond this point the theory becomes strongly coupled and meets the singularity. Indeed, the TeV scale physics yields a larger value for $s^2_W$ that is close to $1/4$. The mass $m_{Z'}$ achieved in this regime may be beyond the singularity, which lies in the invalid regime of the model. This indicates that the high energy behavior of the 3-3-1 model should take the $B-L$ gauge symmetry into account, called the 3-3-1-1 model \cite{3311}, since it not only relaxes the $w$ and $m_{Z'}$ bounds \cite{ds3311} but also cures the non-unitarity of the 3-3-1 model \cite{d3311}.        

Integrating $Z'$ out from (\ref{fin2}) and (\ref{fin1}) that couples both quarks and charged leptons, we get the effective interactions, 
\be \mathcal{H}^{\mathrm{eff(b)}}_{\mathrm{FCNC}}=\fr{g^2[(V^*_{qL})_{3i}(V_{qL})_{3j}]}{4m^2_{Z'}}\left(\bar{q}'_{iL}\ga^\mu q'_{jL}\right)\left[\bar{l}\ga_\mu(1+ \ga_5/3)l\right],\ee which are independent of the Landau singularity too, where $l=e,\mu,\tau$. These interactions potentially contribute to rare semileptonic/leptonic meson decays such as $B\rightarrow K^{(*)} l^+l^-$, $B\rightarrow \pi l^+l^-$, $B_{s,d}\rightarrow l^+ l^-$, and so on. Particularly, let us consider the $b\rightarrow s l^+ l^-$ transition as defined by the effective interactions,
\be \mathcal{H}^{\mathrm{eff(b)}}_{\mathrm{FCNC}} \supset -\fr{4G_F}{\sqrt{2}}[(V^*_{dL})_{32}(V_{dL})_{33}](\Delta C_9 Q_9+\Delta C_{10} Q_{10}),\ee 
where the semileptonic operators are \be Q_9=\fr{\al}{4\pi}(\bar{s}_L\ga^\mu b_L)(\bar{l}\ga_\mu l),\hs Q_{10}=\fr{\al}{4\pi}(\bar{s}_L\ga^\mu b_L)(\bar{l}\ga_\mu \ga_5 l). \ee And, the Wilson coefficients induced by the new physics are identified as 
\be \Delta C_9=3 \Delta C_{10}=-\fr{2\pi}{\al}\fr{m^2_W}{ m^2_{Z'}},\ee where we note that $\fr{G_F}{\sqrt{2}}=\fr{g^2}{8m^2_W}$.

The global fits of the Wilson coefficients to the $b\rightarrow s l^+ l^-$ data have been established by several groups~\cite{wilsoncoe}. Some observed deviations may hint towards an interpretation of the new physics. See \cite{flav331,uhaischhiggs,newbsmm,newbsmm1}, for instance, for recent explanations. A strong constraint on $\Delta C_{10}$ might come from $B_s\rightarrow \mu^+\mu^-$ decay \cite{rarebd}. Here, the new physics contribution is demonstrated by the right graph in Fig. \ref{dsmm} by basic $Z'$ boson exchange. Generalizing the results in \cite{newbsmm1}, we obtain the signal strength, 
\be \mu_{B_s\rightarrow \mu^+\mu^-}=\fr{\overline{Br}(B_s\rightarrow \mu^+\mu^-)}{\overline{Br}(B_s\rightarrow \mu^+\mu^-)_\mathrm{SM}}=1+r^2-2r,\ee where $r=\Delta C_{10}/C_{10}$ ($C_{10}=-4.2453$ is the standard model Wilson coefficient) is real and bounded by $0\leq r\leq 0.1$. It leads to \be m_{Z'}\geq 2.02\ \mathrm{TeV}.\ee Correspondingly, our model predicts $\Delta C_{9}=3\Delta C_{10}=[-1.273,0]$, in agreement with the model-independent global fits \cite{wilsoncoe}. Comparing to the $B_s$ mixing bound, $m_{Z'}\geq 4.67$ TeV, the model prefers the narrower regions of the Wilson coefficient deviations and $r$.  

Note that the $\Delta C_9/\Delta C_{10}$ relation and $m_{Z'}$ bound achieved above agree with the most literature in \cite{flav331} except for the work R. Gauld {\it et al.} Here the contradiction is that the $\bar{l}lZ'$ couplings in the minimal 3-3-1 model like ours (cf. Table II) differ from those used by the mentioned work, because the $l_R$ fields included in the lepton triplets have the $SU(3)_L\otimes U(1)_X$ charges differing from those of which the $l_R$ fields are treated as $SU(3)_L$ singlets\footnote{Although $\beta=-\sqrt{3}$ applies for both cases, i.e. the minimal 3-3-1 model and the model with exotic leptons in triplets/antitriplets, that governs the most gauge couplings.}.                

\subsection{Radiative $\beta$ decay involving $Z'$ as a source of CKM unitarity violation}

CKM unitarity states that $\sum_k V^*_{ik}V_{jk}=\delta_{ij}$ and $\sum_i V^*_{ik}V_{il}=\delta_{kl}$, where we relabel $V=V_{\mathrm{CKM}}$, $i,j=u,c,t$, and $k,l=d,s,b$ for brevity.

The standard model prediction is in good agreement with the above relations \cite{pdg}. However, a possible deviation would be the sign for the violation of CKM unitarity. Considering the first row, the experiments constrain \cite{pdg}
\be \Delta_{\mathrm{CKM}}=1-\sum_{k=d,s,b}|V_{u k}|^2<10^{-3}.\ee This deviation would bound $Z'$ as a result of its loop effects that along with $W$ boson contributes to quark $\beta$-decay amplitudes from which $V_{uk}$ are extracted, including muon decay that normalizes the quark amplitudes. 

Generalizing the result in \cite{betazp}, we obtain 
\be \Delta_{\mathrm{CKM}}\simeq -\fr{3}{4\pi^2}\fr{m^2_W}{m^2_{Z'}}\ln\left(\fr{m^2_W}{m^2_{Z'}}\right)(A^{Z'}_{l_L})_{11}\left\{(A^{Z'}_{l_L})_{11}-\fr 1 2 \left[(A^{Z'}_{u_L})_{11}+(A^{Z'}_{d_L})_{11}\right]\right\},\ee where the fermion couplings are given in the basis of physical left-chiral fields $\mathcal{L}\supset \bar{f}_L\ga^\mu A^{Z'}_{f_L}f_L Z'_\mu$ so that $A^{Z'}_{f_L}\equiv -\fr{g}{c_W}V^\dagger_{fL}a^{Z'}_L(f)V_{fL}$, where $a^{Z'}_L(f)=\fr 1 2 [g_V^{Z'}(f)+g_A^{Z'}(f)]$.    

We have 
\bea &&(A^{Z'}_{l_L})_{11}=-\fr{g}{c_W}a^{Z'}_L(e_a)=-\fr{g}{2c_W}\sqrt{\fr{1-4s^2_W}{3}},\\ &&(A^{Z'}_{u_L})_{11}=-\fr{g}{c_W}a^{Z'}_L(u)=\fr{g}{2c_W}\fr{c_{2W}}{\sqrt{3(1-4s^2_W)}},\\
&&(A^{Z'}_{d_L})_{11}=-\fr{g}{c_W}\left\{a^{Z'}_L(d)+[a^{Z'}_L(b)-a^{Z'}_L(d)]|V_{td}|^2\right\}\simeq \fr{g}{2c_W}\fr{c_{2W}}{\sqrt{3(1-4s^2_W)}}.\eea   
Hence,
\be \Delta_{\mathrm{CKM}}\simeq -\fr{g^2(1-3s^2_W)}{8\pi^2c^2_W}\fr{m^2_W}{m^2_{Z'}}\ln\left(\fr{m^2_W}{m^2_{Z'}}\right)\simeq -0.00215\fr{m^2_W}{m^2_{Z'}}\ln\left(\fr{m^2_W}{m^2_{Z'}}\right).\ee Since $m_W\simeq 80.4$ GeV and $m_{Z'}$ in the TeV range (in fact, $m_{Z'}>4.67$ TeV), we have $\Delta_{\mathrm{CKM}}<10^{-5}$. The effect of CKM unitarity violation due to $Z'$ is negligible and thus the model easily evades the experimental bound. This conclusion contradicts a study of the minimal 3-3-1 model in \cite{flav331}.  

\subsection{LEPII search for $Z'$}

LEPII searches for a neutral gauge boson $Z'$ via the channel $e^+e^-\rightarrow f\bar{f}$ where $f$ is some ordinary fermion \cite{lepii1}. Considering the final state $f\neq e$, the process is governed by the $s$-channel exchange of $Z'$, yielding the effective Lagrangian,
\be \mathcal{L}_{\mathrm{eff}}\supset \fr{g^2}{c^2_W m^2_{Z'}}[\bar{e}\ga^\mu(a^{Z'}_L(e)P_L+a^{Z'}_R(e)P_R )e][\bar{f}\ga^\mu(a^{Z'}_L(f)P_L+a^{Z'}_R(f)P_R)f],\ee where $a^{Z'}_R(f)=\fr 1 2 [g_V^{Z'}(f)-g_A^{Z'}(f)]$, while the left chiral couple has been defined before. 

We are interested in a particular process for $f=\mu$ and the effective Lagrangian becomes 
\be \mathcal{L}_{\mathrm{eff}}\supset \fr{g^2[a^{Z'}_L(e)]^2}{c^2_W m^2_{Z'}}(\bar{e}\ga^\mu P_L e)(\bar{\mu}\ga^\mu P_L\mu)+(LR)+(RL)+(RR),\ee where the latter terms differ from the first term only in chiral structures. 

LEPII studied such chiral interactions and gave respective constraints on the chiral couplings, which are typically in a few TeV \cite{lepii1}. Choosing a typical bound derived for a new $U(1)$ gauge boson like ours, it yields \cite{lepii2}
\be \fr{g^2[a^{Z'}_L(e)]^2}{c^2_W m^2_{Z'}} < \fr{1}{(6\ \mathrm{TeV})^2}.\ee This translates to 
\be m_Z'>\fr{6g}{c_W}a^{Z'}_L(e)\ \mathrm{TeV}=\fr{g}{c_W}\sqrt{3(1-4s^2_W)}\ \mathrm{TeV}\simeq 354\ \mathrm{GeV}.\ee In fact, the $Z'$ mass is in the TeV range, it easily evades the LEPII searches.

\subsection{\label{newphysics} LHC searches for new particle signatures}

\subsubsection{Dilepton and dijet searches}

Because the neutral gauge boson $Z'$ directly couples to quarks and leptons, the new physics process $pp \to l\bar{l}$ for $l=e,\mu$ happens, which is dominantly contributed by the $s$-channel exchange of $Z'$. Among various final states, the dilepton product is worth exploring at the LHC since it has well-understood backgrounds and for models that have the couplings to leptons like ours \cite{lhczp}.  

The cross section for creating a $Z'$ at the LHC and then decaying into a lepton pair can be calculated using the narrow width approximation \cite{lhczp1}
\be \sigma(pp\to Z'\to l\bar{l})= \left[\fr 1 3\sum_{q}\fr{dL_{q\bar{q}}}{dm^2_{Z'}}\hat{\sigma}(q\bar{q}\to Z')\right]\times \mathrm{Br}(Z'\to l\bar{l}).\ee We use the result in \cite{lhczp2} for the parton luminosities $dL_{q\bar{q}}/dm^2_{Z'}$ at the LHC $\sqrt{s}=13$ TeV. The partonic peak cross section $\hat{\sigma}(q\bar{q}\to Z')$ and the branching ratio $\mathrm{Br}(Z'\to l\bar{l})=\Ga(Z'\to l\bar{l})/\Ga_{Z'}$ where $\Ga_{Z'}$ is the total $Z'$ width are obtained as 
\bea
&&\hat{\sigma}(q\bar{q}\to Z')=\fr{\pi g^2}{12c^2_W}[(g^{Z'}_V(q))^2+(g^{Z'}_A(q))^2],\\
&& \Ga(Z'\to l\bar{l})=\fr{g^2m_{Z'}}{48\pi c^2_W}[(g^{Z'}_V(l))^2+(g^{Z'}_A(l))^2],\\
&& \Ga_{Z'}=\fr{g^2m_{Z'}}{48\pi c^2_W}\sum_{f} N_C [(g^{Z'}_V(f))^2+(g^{Z'}_A(f))^2],\eea where we assume the final states in ${Z'}$ decay to be only fermions $f$ which sum over ordinary leptons and quarks, and $N_C$ is the corresponding color number. It is not hard to show that the decay of $Z'$ to ordinary Higgs and/or gauge bosons is subleading. 

\begin{figure}[h]
\bc
\includegraphics[scale=0.8]{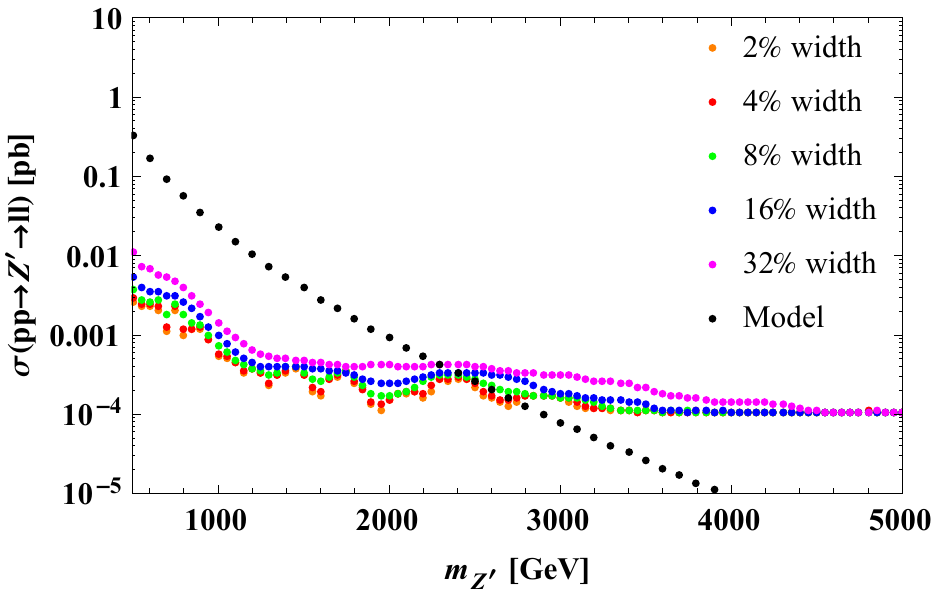}
\caption[]{\label{atlasdilepton} Cross section $\sigma(pp\to Z'\to l\bar{l})$ as a function of $Z'$ mass. Here the colored dots correspond to observed limits for the different widths extracted at dilepton resonance invariant-mass using $36.1\ \mathrm{fb}^{-1}$ of $pp$ collision data at $\sqrt{s}=13$ TeV by the ATLAS detector \cite{lhczp3}. The black dots (almost separated from the others) are the theoretical prediction.}
\ec
\end{figure} 
We show the cross section for the process $pp\to Z'\to l\bar{l}$ in Fig. \ref{atlasdilepton} where $l$ is either electron or muon which has the same $Z'$ coupling. The experimental searches use $36.1\ \mathrm{fb}^{-1}$ of $pp$ collision data at $\sqrt{s}=13$ TeV by the ATLAS collaboration \cite{lhczp3}, yielding negative signal for new high mass events in the dilepton final state. This translates to the lower bound on $Z'$ mass, $m_{Z'}>2.75$ TeV, for the considering model, in agreement with a highest invariant mass of dilepton measured by the ATLAS. 

However, since the $B_s$ mixing hints $m_{Z'}>4.67$ TeV near the Landau singularity, this model indicates an infinitesimal signal strength for dilepton comparing to the current limit as seen from the figure. 

On the other hand, due to such high $Z'$ mass, the dijet bounds~\cite{lhcdijet1} might be relaxed as the above Drell-Yan process is (see also \cite{dydj}). Indeed, the dijet signal strength can be obtained by comparing the couplings $Z'\bar{l}l$ and $Z'\bar{j}j$, which leads to $\sigma(pp\to Z'\to jj)\sim 10^2\sigma(pp\to Z'\to ll)<0.1$ fb for $m_{Z'}>4.67$ GeV. In sort, the new dijet and dilepton signals are negligible if the model obeys the neutral meson mixings.

\subsubsection{Diboson and diphoton searches}

The history of diboson and diphoton searches was fascinated. The ATLAS collaboration at $\sqrt{s}=8$ TeV reported narrow resonances around 2~TeV with high significances in search for new massive dibosons decaying to $WZ$, $WW$, and $ZZ$, respectively \cite{atlas1}. On the other hand, both the ATLAS and CMS collaborations at $\sqrt{s}=13$ TeV indicated a new resonance around 750 GeV decaying to two photons, with a high local significance~\cite{atlascms}. It was later justified that all such resonances were only statistical fluctuations. The latest measurements on diboson and diphoton signatures with invariant masses up to 4-5 TeV reveals no significant excess observed from the standard model background \cite{newdib1,newdib2,newdib3,newdib4}. Additionally, the production cross section times branching ratio of the signatures is set at 95\% CL to be generally below $\mathcal{O}(1)$ fb (diboson) and $\mathcal{O}(0.1)$ fb (diphoton) for the whole range of invariant masses.   

As discussed above, the $Z'$ boson is more massive, i.e. $m_{Z'}>4.67$ TeV, which can account for the ATLAS and CMS diboson and diphoton constraints. Examining various 3-3-1 models, the work in \cite{diboson331} indicates a similar conclusion. Note that $Z'$ does not couple to $WW$ in the effective limit $w \gg u$. Additionally, the $Z'Zh$ coupling is proportional to $10^{-2} u$ and implies a decay width,
\be \Ga(Z'\to Zh)=\fr{\al (1-4s^2_W)}{144 s^2_W c^2_W}m_{Z'}\simeq 0.1\Ga(Z'\to l\bar{l}).\ee Hence, the signal strength, i.e. the cross section times branching ratio, is obtained by \be \sigma(pp\to Z'\to SM\ gauge\ bosons)= 0.1\sigma(pp\to Z'\to l\bar{l}).\ee Comparing to that in Fig. \ref{atlasdilepton}, the predicted diboson signal strength is below 1 fb for invariant mass in the TeV range, in agreement with the latest data \cite{newdib1,newdib2}. 

The new neutral Higgs boson $H$ has the mass $m_H=\sqrt{2\la_2}w$, which can vary from (i.e., just above) the weak scale to TeV scale, depending on the $\la_2$ size. For example, taking $\la_2=0.01$--0.5 and $w=3.9$ TeV due to the FCNC constraint, it leads to $m_H=465$--4000 GeV. Can the new Higgs $H$ account for a significant diphoton excess? First of all, observe that these new Higgs particles can be generated at the LHC that dominantly come from gluon-gluon fusion via exotic-quarks loops, and then partially decay into two high-energy photons\footnote{While this work was in progress, Ref. \cite{diphoton331} reported on the possibility that some third component of scalar triplets could interpret diphoton searches, but their scalar candidates were different. The second article interpreted the diphoton similarly to $H$ but in the minimal 3-3-1 model, whereas the first article considered an extra scalar triplet that did not have VEV but coupled to exotic quarks in the 3-3-1 model with right-handed neutrinos. Ref. \cite{diphoton331beta} did with the 3-3-1 model with arbitrary $\beta$.}. Using the relation $-h^J/\sqrt{2}m^J=1/w$ due to (\ref{exmass}) and the number of exotic quarks $N_J=3$, the $H$ coupling to $G_i$ gluons induced by $J_a$ loops takes the form \cite{newbook} \be \mathcal{L}_{HGG}=\fr{\al_s}{12\pi}\fr{3}{w}HG_{i\mu\nu}G^{\mu\nu}_i,\label{hggdd} \ee which is enhanced due to the number of heavy quarks and is obviously independent of the $h^J$ strength. Because the Higgs mass is proportional to the new physics scale through $m^2_H/w^2= 2\la_2$, the production cross section obeys \be \sigma(pp\rightarrow H)\simeq 0.56\la_2\ \mathrm{pb},\ee which is obtained at the LHC $\sqrt{s}=13$ TeV and using the gluon luminosity \cite{lhczp2}. Hence, $\sigma(pp\rightarrow H)\simeq 5.6$ fb--0.28 pb according to $\la_2$ in the range of interest (correspondingly for the range of $m_H$ by fixing $w=3.9$ TeV). 

If the $H$ boson decays only into two photons, photon and $Z$, and two gluons via the exotic quarks loops, we have \bea &&\fr{\Ga(H\rightarrow \gamma\gamma)}{\Ga(H\rightarrow \gamma Z)}\simeq \fr{1}{2 t^2_W (1-m^2_Z/m^2_H)^3}\simeq 1.74,\\ 
&&\fr{\Ga(H\rightarrow \gamma\gamma)}{\Ga(H\rightarrow GG)}\simeq \fr{19^2}{18}\left(\fr{\al}{\al_s}\right)^2\simeq 8.36\times 10^{-2}.\eea Hence, the gluon mode $H\to GG$ dominates, while the two others $H\to \ga \ga,\ga Z$ are comparable and more enhanced than those of the 3-3-1 model with right-handed neutrinos (where $19^2/18$ changes to $2/9$) due to the large electric charges of exotic quarks. Correspondingly, we have $\mathrm{Br}(H\rightarrow \gamma\gamma)\simeq 7.38\times 10^{-2}$ and $\mathrm{Br}(H\rightarrow \gamma Z)\simeq 4.24\times 10^{-2}$. 

The model predicts \bea && \sigma_{\gamma\gamma}\equiv \sigma(pp\rightarrow H)\mathrm{Br}(H\rightarrow\gamma\gamma)\simeq 0.413\la_2\ \mathrm{fb}\simeq 0.0041\mbox{--}0.20\ \mathrm{fb},\\ && \sigma_{\gamma Z}\equiv \sigma(pp\rightarrow H)\mathrm{Br}(H\rightarrow\gamma Z)\simeq 0.237\la_2\ \mathrm{fb}\simeq 0.0023\mbox{--}0.11\ \mathrm{fb},\eea at $\sqrt{s}=13$ TeV for $m_H=465$--4000 GeV, respectively. The cross section $\sigma_{\gamma\gamma}$ is in good agreement with the measurement in the  regime of measured diphoton invariant mass \cite{newdib3,newdib4}. On the other hand, the 3-3-1 model with right handed neutrinos would predict lower signal strengths for $pp\to H\to \ga\ga,\ga Z$ \cite{diphoton331}. 

Since the natural mass regime of $H$ is radically higher than the masses of the standard model heavy particles, $W$, $Z$, $h$, and $t$, the above signal strengths can be reduced due to the corresponding decay channels as opened. This is because $H$ might couple and decay to $WW$, $ZZ$, and $tt$ due to the $h$-$H$ mixing set by the strength $\xi\simeq (\la_3u)/(\la_2w)$ and to $hh$ due to the scalar coupling $V\supset \fr{\la_3}{2}w H h^2$. Let us evaluate, for example, \bea &&\Ga(H\rightarrow WW)/\Ga(H\rightarrow GG)\simeq 459.5\times (0.1/\al_s)^2 (\la_3/\la_2)^2,\\ 
&&\Ga(H\rightarrow hh)/\Ga(H\rightarrow GG)\simeq 58.16\times (0.1/\al_s)^2 (\la_3/\la_2)^2.\eea Consequently, all these modes $H\to WW,hh$, and so forth may decrease $\mathrm{Br}(H\rightarrow \gamma\gamma)$ and thus $\sigma_{\ga\ga}$ substantially, depending on $\la_3/\la_2$, which can totally fit a future sensitive (low) signal for the process.     

Otherwise, we interpret the inert scalar $H'_3$ or $A'_3$ by omitting the corresponding $Z_2$ symmetry as the diphoton signal instead of $H$, where $H'_3$ and $A'_3$ are the real and imaginary parts of the third component of the inert scalar triplet $\phi=\chi'$ \cite{dns}. Here  we still have $\phi=\eta',\ \sigma$ for dark matter. The $\chi'$ inert triplet has the gauge quantum numbers similarly to $\chi$ but cannot develop VEV due to the potential minimization conditions as supplied in \cite{dns}. This triplet can now couple to the exotic quarks via the Lagrangian $\mathcal{L}_{Y}\supset h'^{J}_{33}\bar{Q}_{3L}\chi' J_{3R}+h'^{J}_{\al\beta}\bar{Q}_{\al L}\chi'^* J_{\beta R}+H.c.$. The $H'_3J_aJ_a$ couplings and $J_a$ masses (as induced by $H$) are not correlated. Because of $\langle \chi'\rangle=0$, there is no $H'_3$-$h$ mixing and no $H'_3hh$ coupling, as expected (similarly valid for $A'_3$, but it will be skipped hereafter). The $H'_3$ coupling to gluons is induced by $J_a$ loops which yields the effective Lagrangian, 
\be \mathcal{L}_{H'_3 GG}=\fr{\al_s}{12\pi}\fr{h'^J}{h^J}\fr{3}{w} H'_3 G_{i\mu\nu} G^{\mu\nu}_i.\ee Taking $m_{H'_3}$ in the actual range of diphoton search, we have \be \sigma(pp\rightarrow H'_3) \simeq 0.28\left(\fr{h'^J}{h^J}\right)^2 \left(\fr{m_{H'}}{w}\right)^2\ \mathrm{pb},\ee at the LHC $\sqrt{s}=13$ TeV. Note that in this case $H'_3$ dominantly decays to gluons, which leads to $\mathrm{Br}(H'_3\rightarrow \gamma\gamma)\simeq 7.38\times 10^{-2}$ and $\mathrm{Br}(H'_3\rightarrow \gamma Z)\simeq 4.24\times 10^{-2}$, both of which are induced by $J_a$ loops, analogous to the above case. Correspondingly, the model predicts \bea && \sigma(pp\rightarrow H'_3)\mathrm{Br}(H'_3\rightarrow \gamma\gamma)\simeq 0.2(h'^J/h^J)^2(m_{H'}/w)^2\ \mathrm{fb},\\ 
&&\sigma(pp\rightarrow H'_3)\mathrm{Br}(H'_3\rightarrow \gamma Z)\simeq 0.118(h'^J/h^J)^2(m_{H'}/w)^2\ \mathrm{fb}.\eea The first one fits the current bound yielding the diphoton mass proportional to \be m_{H'}\sim w h^J/h^{J'},\ee which does not require a large hierarchy in the Yukawa couplings for the dilepton mass in the TeV range and $w=3.9$--5 TeV, unlike \cite{diphoton331}.

Although the couplings of $H'_3$ to $tt$, $WW$, $ZZ$, and $hh$ are suppressed at the tree level, they might still decay to these channels as induced by loops, e.g. to $tt$ by a triangular loop of two $J_3$ and a new gauge boson $X_\mu$ due to the gauge interactions $XtJ_3$ as in (\ref{cccd}) and to $ZZ$ by $J_a$ loops due to the gauge interactions $ZJ_aJ_a$ as in Table \ref{tab1}. But, these radiative decays are all rare since the invariant amplitudes are suppressed by the loop factor $1/16\pi^2$ and the new physics scale $\sim u/w$. For instance, assuming $m_J=m_X$ for brevity, the effective coupling $-\fr{m_t}{u}c_{\mathrm{eff}} H'_3 \bar{t}t$ is obtained by $c_{\mathrm{eff}}=\fr{g^2}{96\pi^2}\fr{u}{w}$, which is negligible, as expected.                         

Last, but not least, $H$ can mainly decay to two electroweak gauge bosons, but it can easily evade the ATLAS and CMS diboson bounds. This is because, for instance, at the LHC $\sqrt{s}=8$ TeV and for $m_H=2$ TeV (i.e. $\la_2 = 1/8$) and $w=3.9$ TeV, we have $\sigma(pp\rightarrow H)\simeq 4.68\times 10^{-2}$ fb as induced by $J_a$ loops, which yields a too small signal strength, in agreement to the data. Also, the inert scalars such as $H'_3$ and $A'_3$ could easily evade the diboson bound. Although some of them can be enough produced by choosing appropriate $h'^J/h^J$ ratio, its decays into the electroweak bosons are very rare due to the loop factor and $u/w$ suppressions, aforementioned.        

In summary, the previous results of the diphoton and diphoton searches are refined. Additionally, an alternative solution to the diphopton signal as interpreted in \cite{diphoton331} is obtained by the simple 3-3-1 model with inert scalars.

\subsubsection{Monojet and dijet dark matter}

The model possesses potential dark matter candidates, a singlet scalar in $\chi'$, a doublet scalar in $\eta'$, and a triplet scalar in $\sigma$, in which their relic density and direct and indirect searches were studied in \cite{dns}. In what follows, we consider $\eta'$ signatures at the LHC. The other cases could be similarly done. 

Let us write explicitly \bea \eta'=
\left(
\begin{array}{c}
\fr{1}{\sqrt{2}}(H'_1+i A'_1\\
H'^-_2\\
H'^+_3
\end{array}\right).\eea Here we interpret the dark matter candidate as $H'_1$, without loss of generality. The dark field masses are not degenerate. The abundance and (in)direct detection imply that these masses are either in a low regime above the weak scale or in the TeV regime \cite{dns}. A resonant regime set by $m_{H'_1}=\fr 1 2 m_{H}$ can be viable in the middle scale for $m_H$ just above 1 TeV. 

At the LHC, the dark matter candidate can be directly produced, realized in form of a large missing transverse momentum or energy. Desirable experimental signal would be an excess of a mono-$X$ or two $X$'s final state, recoiling against such missing energy. Here $X$ may be a jet, a lepton, electroweak gauge/Higgs bosons, and even new particles. 

If $H'_1$ has a mass in the low regime, it can be created via exchanges of the standard model particles $W$, $Z$, and $h$, similarly to the inert doublet model. This model has been extensively studied \cite{idmlhcsearch} and the relevant results can apply to our model, since the other new physics is mostly decoupled. We are interested in the TeV mass regime of dark matter that explicitly probes the simple 3-3-1 model with inert scalars.

\begin{figure}
\bc
\includegraphics[scale=0.7]{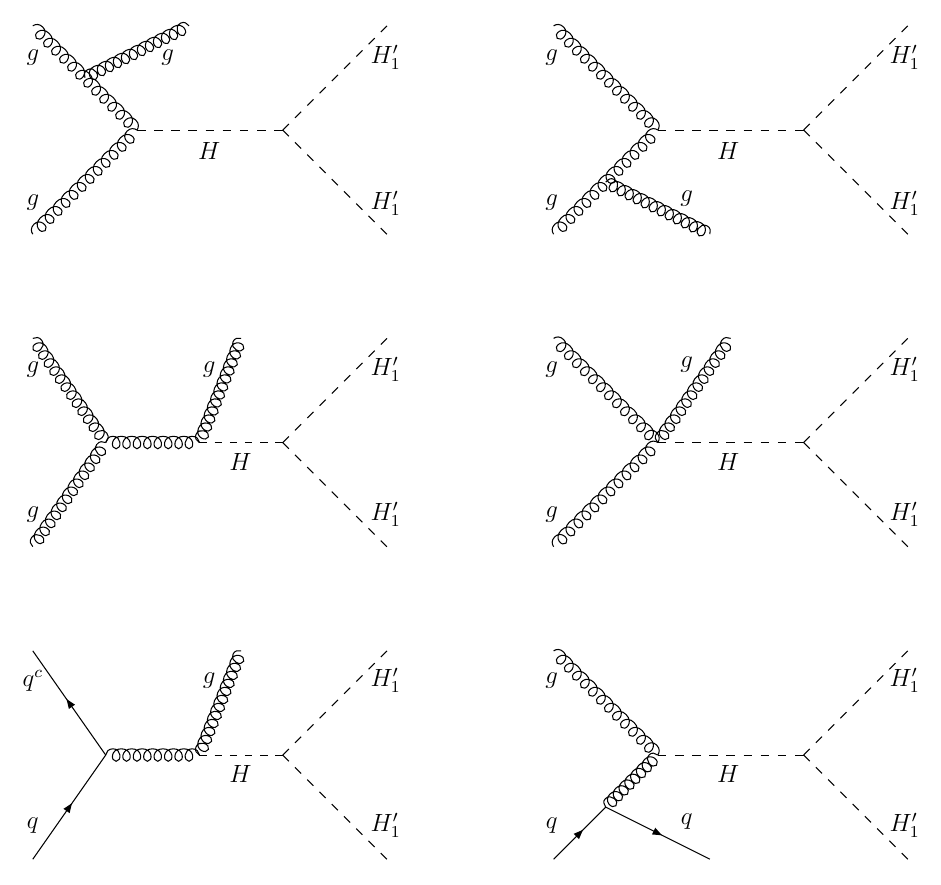}
\includegraphics[scale=0.7]{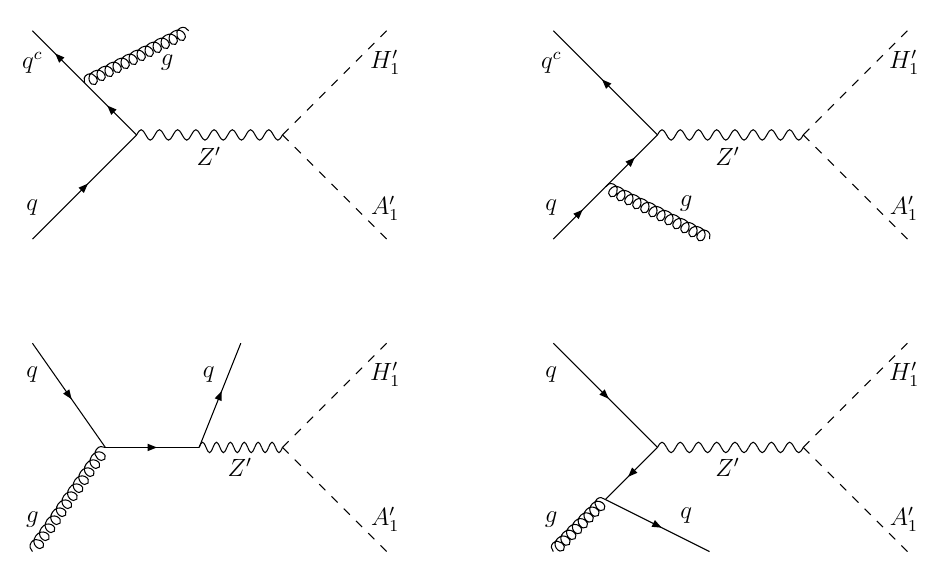}
\caption{\label{monojetdm} Monojet production processes associated with a pair of dark matter.}
\ec
\end{figure}

Dark matter and normal matter can couple via new Higgs and gauge portals governed by $H$ and $Z'$ due to the interactions $HH'_1H'_1$ and $Z'H'_1A'_1$ as well as those of $H,Z'$ to normal particles. Although the gauge boson $Z'$ obtain a high mass, it tree-level couples to partons (i.e. quarks) unlike $H$ which is relevant to this search. That said, both of the portals might contribute equivalently to the monojet production at the LHC. Therefore, the mono-$X$ signature includes dominantly a jet via the $H$-exchanged processes $gg\to g H'_1H'_1$, $qq^c\to g H'_1H'_1$, and $gq\to q H'_1H'_1$ as well as via the $Z'$-exchanged processes $qq^c\to g H'_1A'_1$ and $gq\to q H'_1A'_1$. The Feynman diagrams are given in Fig. \ref{monojetdm}. Let us recall that the $H$ boson interacts with gluons via $J_a$ loops as given in~ (\ref{hggdd}).

Note that $H$ and $Z'$ couple to $H'_1$ and $A'_1$ as \cite{dns} 
\be \mathcal{L}\supset -\fr 1 2 x_3 w H H'_1 H'_1 + \fr{g}{2}\sqrt{\fr 1 3 -t^2_W}H'_1\overleftrightarrow{\pa}^\mu A'_1 Z'_\mu.\ee
Since $H,Z'$ are heavy, the above diagrams can be generated by the effective (contact) interactions after integrating them out, 
\be \mathcal{L}_{\mathrm{eff}}=-\fr{x_3\al_s}{8\pi m^2_H}G_{i\mu\nu}G_i^{\mu\nu}H'^2_1+\fr{g^2\sqrt{1-4s^2_W}}{4\sqrt{3}c^2_W m^2_{Z'}}\bar{q}\ga_\mu[g^{Z'}_V(q)-g^{Z'}_A(q)\ga_5]q H'_1\overleftrightarrow{\pa}^\mu A'_1.\ee Of course, the number of monojet diagrams generated by these interactions correspondingly matches those in Fig. \ref{monojetdm}. The monojet production via the effective interactions---which appropriately work for the case of a large mass mediator with a small transmitting momentum---were studied in the literature; see for instance \cite{monojeteffective}. The missing transverse energy at the LHC for $\sqrt{s}=13$~TeV is below $\mathcal{O}(1)$ TeV \cite{monojetlhc13} which is radically smaller than $m_{H,Z'}$ and can be applied to our case. 

Generalizing the result in~\cite{fmonojeteff}, the strongest bounds come from the effective interaction with $H$ exchange and the strongest one with $Z'$ exchange according to the largest gauge coupling, namely $g_{\mathrm{max}}\equiv \mathrm{Max}\{g^{Z'}_V(q),g^{Z'}_A(q)\}=(1+2s^2_W)/2\sqrt{3(1-4s^2_W)}$, which yield 
\be \fr{x_3\al_s}{8\pi m^2_H} < \fr{1}{(3\ \mathrm{TeV})^2},\label{dadd1}\ee
 \be \fr{g^2\sqrt{1-4s^2_W}}{4\sqrt{3}c^2_W m^2_{Z'}}g_{\mathrm{max}} < \fr{1}{(0.3\ \mathrm{TeV})^2}.\label{dadd2}\ee The condition (\ref{dadd1}) translates to $m_H>200\sqrt{x_3}\simeq 20,\ 200$, and 735 GeV for $x_3=0.01,\ 1$, and $4\pi$, respectively. Here the first value $x_3=0.01$ was taken from \cite{dns} once constraining the relic density and (in)direct searches, while the last value approaches the perturbative limit $x_3=4\pi$. The condition (\ref{dadd2}) implies $m_{Z'}>55$ GeV for $s^2_W=0.231$ and $\al=1/128$. The mediator masses required are small due to the weakness of the couplings. 
 
Conversely, since $H,Z'$ are indeed in the TeV range, the monojet signal associated with dark matter pair production in this model is radically smaller than the current bound. A future sensitivity project is needed to prove or rule out our model.          
                                                       
\section{\label{conclusion} Conclusion}

The baryon minus lepton number ($B-L$) is an accident symmetry in the standard model which always commutes with the gauge symmetry. It can even act as a hidden gauge symmetry, if three right-handed neutrinos are included. The case is different and explicit in the 3-3-1 extensions, in which the $B-L$ and 3-3-1 symmetries neither commute nor close algebraically, by contrast. Consequently, one must either work with a more fundamental gauge group that encloses those algebras due to consistency, known as the 3-3-1-1 gauge symmetry \cite{3311,ds3311,d3311}, or accept the 3-3-1 models but in this case $B-L$ is an approximate symmetry and the unitarity is not ensured at a high energy scale \cite{dpsadd1,dns}. 

Interpreting inert scalar multiplets in the 3-3-1 models is a natural recognition of the second fold because it yields not only appropriate $B-L$ violating interactions that make the model viable, but also providing novel phenomenology in neutrino mass generation and dark matter \cite{dpsadd1,dns}, as well as the global fit of $\rho$-parameter and possible solutions to the LHC new physics anomalies. Here the last two points are obtained in the current work. The 3-3-1 model considered in this work contains the most minimal lepton and normal-scalar contents, while its inert sector includes $\eta'$ and $\chi'$ as replications of the normal scalars, as well as the scalar sextet $\sigma'$, that have the definite roles. 

Because of such a special scalar sector, the model is calculable. All the gauge interactions of fermions and scalars are derived and the self-interactions of scalars as well as the Yukawa interactions are achieved. The standard model interactions are recovered in the effective limit $u\ll w$. It is noted that the interactions of charged leptons with $Z'$ are not governed by the general formulae for $g_{V,A}$ since the right-handed charged leptons possess the nonzero $T_8$ charge. With those at hand, we obtain the production cross-section and branching ratios of the standard model like Higgs boson, the effective interactions describing the neutral meson mixings and rare semileptonic/leptonic meson decays, the radiative beta decay, as well as the collision processes, due to the $Z'$ and other new contributions. 

The new physics contributions to the standard model Higgs signal strength are strictly suppressed by $(u/w)^2$, and in the viable parameter regime the model's prediction is close to the standard model value, in good agreement with the data. The $B_s-\bar{B}_s$ mixing places the strongest  bound on the 3-3-1 breaking scale such as $w>3.9$ TeV, while the data of rare $B_s\rightarrow \mu^+\mu^-$ decay indicate $m_{Z'}>2.02$ TeV. The effect of CKM unitarity violation and the signal of $Z'$ at the LEPII are all small, compared to the current bounds. Our model predicts the new physics contribution to the Wilson coefficients to be $\Delta C_9=3\Delta C_{10}=[-1.273,0]$, in agreement with the global fits. Such results improve a previous study supplied in \cite{flav331} since the $Z'\bar{l}l$ coupling is changed particularly for the minimal 3-3-1 model like our model. The lower bound on $w$ from the meson mixing can be relaxed due to the $B-L$ gauge interaction once included and its kinetic mixing with $Z'$, where both the new neutral gauge bosons substantially contribute and modify the bound \cite{ds3311}.    

We have shown that $Z'$ can be produced at the LHC and decay to dilepton $l=e,\mu$ which reveals a bound $m_{Z'}>2.75$ TeV, smaller than that from the meson mixing. With the largest bound on $m_{Z'}$, the dilepton, dijet, and diboson signals due to $Z'$ decays are small, compared to the current limits. Additionally, the new neutral Higgs boson $H$ can be created and decay to diboson at the LHC but the signal is small too. It is noteworthy that $H$ can decay to diphoton at the LHC, in good agreement to the experiment. Also, the inert scalars can address the current bound of the diphoton signal, while they can easily evade the diboson signal constraint, since their production cross-sections are too small in comparison to the data. Let us note that the inert scalar $H'_3$ or the inert pseudoscalar $A'_3$ may be naturally interpreted as the diphoton signal in our model, since their decay modes to $\gamma\gamma$ are more enhanced due to the large electric charges carried by the exotic quarks; hence the hierarchy in the exotic quark Yukawa couplings is not necessary. The model yields $\sigma(pp\rightarrow H'_3)Br(H'_3\rightarrow \gamma Z)$ at the LHC comparable to (slightly smaller than) the $\gamma\gamma$ production strength. It is emphasized that the exotic quarks and inert scalars are fundamental components of a previously-studied 3-3-1 model. We have examined the monojet signal associated with a pair of dark matter $H'_1$ presumably produced at the LHC. The experiments limit the portals $H$ and $Z'$ to be light, that opposes the other bounds. In other words, the monojet signature is negligible, below the detection limit due to the TeV-scale masses of $H,Z'$.        

The results as obtained obviously reveal an significant inert scalar sector, strongly correlated to the simple 3-3-1 model on both the theoretical and phenomenological sides. The essential contacts between the two sectors are (1) the mathematical inconsistency of $B-L$ symmetry is cured, (2) dark matter and other new particle signatures are explained, (3) the $\rho$-parameter agrees with the global fit, and (4) the small neutrino masses recognize an approximate $B-L$ symmetry.                                                                

\section*{Acknowledgments}

This research is funded by Vietnam National Foundation for Science and Technology Development (NAFOSTED)
under grant number 103.01-2016.77. P.V.D. would like to thank Dr. C. A. de S. Pires for a comment on the new-physics scale due to the $\rho$-parameter constraint, which has been solved in the present work, and Dr. Do Thi Huong for a computational code of the Drell-Yan process. N.T.K.N. expresses gratitude to Dr. Dang Van Soa for his guidances of the 3-3-1 model.

\end{document}